\documentclass[showpacs,preprintnumbers,amsmath,amssymb,prb,twocolumn,floatfix,amssymb]{revtex4}
\usepackage[dvips]{graphicx,color}
\usepackage{dcolumn}
\usepackage{amsmath,amssymb,bbold,bm}

\begin{document}
\title{Charge-Kondo Effect in Mesoscopic Superconductors Coupled to Normal Metals}
\author{Ion Garate$^{1,2}$}
\affiliation{$^1$Department of Physics and Astronomy, The University of British Columbia, Vancouver, BC V6T 1Z1, Canada}
\affiliation{$^2$Canadian Institute for Advanced Research, Toronto, ON M5G 1Z8, Canada.}
\date{\today}

\begin{abstract}
We develop a theoretical proposal for the charge-Kondo effect in mesoscopic normal-superconductor-normal heterostructures, where the superconducting gap exceeds the electrostatic charging energy.
Charge-Kondo correlations in these devices alter the conventional temperature-dependence of Andreev reflection and electron cotunneling.
We predict typical Kondo temperatures of $\gtrsim 10 {\rm mK}$, and suggest experimental architectures that combine superconducting charge-qubits with semiconducting nanowires at cryogenic temperatures.
\end{abstract}
\maketitle

\section{Introduction}

Five decades after its discovery, the Kondo effect is regarded as an archetype many-body phenomenon that interconnects strongly correlated systems such as atomic nuclei, heavy fermion compounds, superconducting cuprates and quantum dot devices.\cite{cox1998}  
The Kondo effect emerges from the interaction of localized, degenerate states with itinerant degrees of freedom.
It is characterized by a low-temperature infrared divergence in perturbative calculations of physical observables such as resistivity and magnetic susceptibility.  
Often the localized degrees of freedom originate from electronic spins and their degeneracy is tied to the usual spin degeneracy.   
The ensuing spin-Kondo effect (SKE) is realized when dilute magnetic impurities are embedded in a non-magnetic host, or when conducting electrodes are coupled to quantum dots.\cite{glazman2005} 

Aside from spin, any degenerate two-level system that is coupled to a fermionic bath is a potential host for the Kondo effect. 
In particular the charge counterpart of SKE, known as charge-Kondo effect (CKE), arises when a Fermi sea hybridizes with dilute impurities containing two degenerate charge states.
This variant has received sustained theoretical attention for the last two decades, being investigated in non-superconducting single-electron devices\cite{matveev1991,glazman1990,matveev1995,zarand2000,buxboim2003,lehur2004,kakashvili2007} as well as in valence-skipping elements with attractive onsite interactions (``negative-U molecules'') .\cite{taraphder1991,cornaglia2004,dzero2005,koch2007,laad2009} 
In the former case the electrostatic energy can be engineered by gate electrodes to be the same for two states that differ by one electron, whereas in the latter case chemistry dictates that the charging energy be degenerate between states that differ by two electrons.
Incidentally, CKE in single-electron devices constitutes a paradigm of the two-channel Kondo effect, where the channels originate from spin.

As opposed to SKE, which has been thoroughly observed both in its single-channel and two-channel versions,\cite{ske exp} the experimental detection of CKE remains at a primitive stage. 
Experimental challenges abound for the implementation of CKE in non-superconducting devices.
On one hand, CKE is sensitive to and washed out by background charge noise, much like SKE is sensitive to and washed out by magnetic fields. 
On the other hand, both the charge-Kondo temperature and the charging energy of the system are required to be large compared to the single-particle energy-level spacing in the grain, which is difficult to achieve in semiconductors.
Prospects may be better for metallic systems, wherein single-particle energy levels form a near-continuum.
Yet metallic devices often involve junctions with multiple channels, and unfortunately the charge-Kondo temperature in non-superconducting systems scales exponentially unfavourably with the number of channels.\cite{zarand2000} 
Some of these problems might be circumvented by using atomic point contacts\cite{zarand2000} or else by resorting to resonant tunneling devices.\cite{gramespacher2000, lebanon2003}
Altogether, attempts to measure fingerprints of the CKE in non-superconducting systems have thus far met with suggestive yet inconclusive outcomes.\cite{berman1999,lehnert2003} 

Recent experimental efforts\cite{negative U experiment} concerning negative-U molecules appear to have met with more success:
charge-Kondo temperatures of $\sim 5 {\rm K}$ have been reported in PbTe doped with valence-skipping Tl.
Unfortunately, real materials with valence-skipping compounds are relatively rare, more complicated and less tunable than artificial single-electron devices.

In this work we theoretically demonstrate that CKE can also occur in artificially fabricated mesoscopic superconducting islands that are connected to non-superconducting leads (NSN devices).
We concentrate on superconducting grains whose many-particle energy-gap ($\Delta$) exceeds the electrostatic charging energy ($E_c$). 
These grains behave somewhat like giant negative-U molecules, and they are to valence-skipping elements what quantum dots are to local magnetic moments: more easily manufacturable and more readily controllable systems, albeit at the price of a parametrically lower Kondo temperature.

Our proposal is a natural and perhaps obvious extension of Refs.~[\onlinecite{matveev1991,glazman1990}] and ~[\onlinecite{koch2007}], as it combines attractive (phonon mediated) and repulsive (electrostatic) Coulomb interactions with a dense single-particle energy spectrum. 
Surprisingly, there have been no thorough studies of CKE in superconductors with $\Delta>E_c$.
The relevant literature is limited to some peripheral statements,\cite{zaikin1994,dubi2007} along with a tacit assumption that CKE in superconducting dots is conceptually similar to CKE in non-superconducting single-electron devices (NNN devices).
Contrary to this view, our results aim to draw the attention of theorists and experimentalists alike towards the study and search of charge-Kondo correlations in NSN systems. 
%In particular, our version of CKE is free from some of the experimental difficulties mentioned above.

The rest of this paper is organized as follows.
In Section~\ref{sec:model} we introduce the basic microscopic model for a NSN heterostructure.
The island is assumed to be chaotic and smaller than the superconducting coherence length, thereby allowing for coherence between electrons tunneling through different junctions. 
In Section~\ref{sec:qimp} we map the superconducting island onto an artificial spin 1/2 describing two charge states.
This mapping is well-known and has already been exploited in existing Cooper pair boxes.
In Section~\ref{sec:kondo} we attach normal metallic leads to the superconducting island, and find that at low enough energies and near the charge degeneracy point this system is described by an anisotropic Kondo model.
The degree of anisotropy and the magnitude of the Kondo couplings can be changed by tuning the ratio $\Delta/E_c$.
In Section~\ref{sec:disc} we provide quantitative estimates for the charge-Kondo temperature in experimentally realizable NSN devices.
In addition, we present a detailed discussion on how our proposal for CKE fares in comparison with previous proposals concerning negative-U molecules and non-superconducting devices. 
%In Section  we comment on the possibility of arriving at a single-channel Kondo model in wide normal-superconductor junctions.
%This is an important possibility, as it allows to reach measurable Kondo temperatures in existing devices.
%In contrast, no such possibility arises in previous proposals of CKE based on non-superconducting single-electron transistors.
In Section~\ref{sec:cond} we calculate the fingerprints of CKE in physical observables of NSN systems. 
Specifically, we focus on the temperature-dependence of the zero-bias conductance at low temperatures.
In Section~\ref{sec:bozo} we determine the fate of CKE when the normal-superconducting junctions are highly transparent; the outcome depends on the strength of electron-electron interactions in the normal metallic leads.
Section~\ref{sec:conc} is devoted to a short summary and conclusions, and the Appendix contains technical calculations that rigorously justify the considerations of Section~\ref{sec:kondo}.

\section{Microscopic Model}
\label{sec:model}

Let us suppose we have a mesoscopic superconducting grain that is weakly coupled to normal metallic leads.
Its Hamiltonian can be expressed as
\begin{eqnarray}
\label{eq:model}
{\cal H}&=&{\cal H}_{\rm l}+{\cal H}_{\rm d}+{\cal H}_T\nonumber\\
{\cal H}_{\rm l} &=& \sum_{\alpha k \sigma} \xi_{k} c^\dagger_{\alpha k \sigma} c_{\alpha k \sigma}\nonumber\\
{\cal H}_{\rm d} &=& \sum_{n \sigma} \epsilon_n d^\dagger_{n\sigma} d_{n\sigma}+E_c(\hat{N}-N_g)^2+\eta\hat{T}^\dagger \hat{T}\nonumber\\
{\cal H}_T &=& \sum_{\alpha k \sigma n} t_{\alpha k n} c^\dagger_{\alpha k \sigma} d_{n \sigma} +{\rm h.c.}.
\end{eqnarray}
${\cal H}_{\rm l}$ is the Hamiltonian of the leads. 
$c^\dagger_{\alpha k \sigma}$ creates an electron with momentum $k$ and spin $\sigma$, in a lead labeled by $\alpha$ ($\alpha={\rm L(eft),R(ight)}$). 
$\xi_{k}$ is the energy dispersion of the itinerant fermions measured from the Fermi energy. 
For now we assume that there is only one conduction channel on each lead; as such, $k$ is the momentum perpendicular to the interface between the normal-metal and the superconductor (NS interface).
Generalizations to multi-channel leads, which are necessary when the linear dimensions of the NS interface exceed the Fermi wavelength, will be discussed in Section V.
Likewise we neglect electron-electron interactions in the leads, although this assumption will be relaxed in Section VIII.

${\cal H}_{\rm d}$ is the Hamiltonian of the superconducting quantum dot. 
$d^\dagger_{n\sigma}$ creates an electron in the $n$-th single-particle level of the island. 
The corresponding energy (measured with respect to the Fermi energy) can be written as $\epsilon_n=n \delta$, where $\delta$ is the average single-particle level spacing in the dot.
Hence, in this model every energy-level is non-degenerate, save for the twofold spin degeneracy. 
$\hat{N}=\sum_{n\sigma} d^\dagger_{n\sigma} d_{n\sigma}$ and $\hat{T}=\sum_n d_{n\downarrow}d_{n\uparrow}$ are the number and pairing operators in the island, respectively.
$\eta$ is the effective (phonon mediated) coupling between electrons in the island.
Hereafter we adopt the BCS approximation\cite{tinkham1996} by introducing an s-wave superconducting gap $\Delta\equiv\eta\langle \hat{T}\rangle$.
The BCS approximation is accurate\cite{delft2001} provided that $\Delta>\delta$, which constitutes the regime of interest in this paper.
For reasons exposed below, we take $\Delta>E_c$, where $E_c=e^2/2 C_\Sigma$ is the charging energy of the island.
 $C_\Sigma=C_L+C_R+C_{\rm g}$ is the total capacitance device, $C_{L(R)}$ is the capacitance of the left (right) NS junction and $C_{\rm g}$ is the gate capacitance.
In mesoscopic NSN heterostructures most of the capacitance emanates from the junctions and scales roughly as $C_\Sigma\sim \epsilon A/d$, where $\epsilon$ is the dielectric constant of a thin insulating layer that separates the normal metal from the superconductor, $A$ is the surface-area of each NS junction and $d\simeq 1{\rm nm}$ is the thickness of the insulating layer.
$N_g$ is the charge induced in the island by the gate electrode.
We neglect intra-dot exchange interactions, which is adequate given the spin-singlet order parameter of the superconductor.

${\cal H}_T$ models the single-particle tunneling between the island and the leads, $t_{\alpha k n}$ being the tunneling amplitude.
Throughout this paper (with the exception of Section VIII) we treat the tunneling term as a weak perturbation. 
Hence we require that $\delta>>\Gamma$, where $\Gamma$ is the broadening of the energy-levels of the island due to tunneling between the island and the leads.

Our model relies on precepts from random matrix theory (RMT) and is valid only for dots without spatial symmetries and for states in the vicinity of the Fermi energy.\cite{aleiner2002,glazman2005}
According to RMT, single-particle states with $|\epsilon_n|\lesssim E_T$ are non-degenerate except for their spin degeneracy, where
$E_T$ is the Thouless energy. 
$E_T=\hbar D/L^2$ for a dirty sample (where $D$ is the diffusion constant and $L$ is the linear dimension of the superconductor) and $E_T=\hbar v_F/L$ for a clean sample (where $v_F$ is the Fermi velocity).
Anticipating the fact that charge-Kondo correlations originate mainly from states with $|\epsilon_n|\lesssim \Delta$, we request $E_T\gtrsim\Delta$ so that Eq.~(\ref{eq:model}) can capture CKE quantitatively.
This condition is tantamount to $L\lesssim \zeta$, where $\zeta$ is the superconducting coherence length. 
In sum, the hierarchy of relevant energy scales reads
\begin{equation}
\label{eq:hierar}
E_T\gtrsim\Delta>E_c>\delta>\Gamma.
\end{equation}
$L\lesssim \zeta$ implies electron coherence between the two junctions, which in turn gives rise to elastic electron cotunneling and crossed Andreev reflection (Section~\ref{sec:cond}).
Unmentioned as it is in Eq.~(\ref{eq:hierar}), we note that the Debye frequency $\omega_D$ (such that $\Delta\neq0$ iff $|\epsilon_n|\lesssim\omega_D$) is much larger than $E_T$ for all cases discussed in this paper.
%Should $L$ exceed $\zeta_0$ significantly, it would be more appropriate to use three dimensional momentum eigenstates for the superconductor.
%This latter scenario is unfavourable for the experimental detection of CKE, as will be discussed in Section VI.

\section{Isolated Superconducting Island: a Pseudospin 1/2 Quantum Impurity}
\label{sec:qimp}

In this section we ignore the leads and simply consider an electrostatically gated superconducting island, which contains a definite number of particles $N$ because $[\hat{N}^2,\hat N]=[\hat{T}^\dagger\hat{T},\hat N]=0$.
Its ground state depends on $N$ as\cite{averin1992} 
\begin{equation}
\label{eq:E_0}
E_0 (N)=E_c(N-N_0-N_g)^2+\frac{\Delta}{2}(1-(-1)^N),
\end{equation}
where $N_0$ is the number of electrons in the island when the gate voltage is zero.
We assume without loss of generality that $N_0$ is even so that $N-N_0$ and $N$ have the same parity; herein we absorb $N_0$ in the definition of $N$.
The last term of Eq.~(\ref{eq:E_0}) imposes an additional energy cost whenever $N$ is odd, because of a single unpaired quasiparticle that must reside outside the superconducting condensate.
The lowest energy this quasiparticle can have equals $\Delta$.
If $\Delta> E_c$ and if the gate voltage is tuned to $N_g=2M+1$ (for any integer $M$), the lowest energy eigenvalues are (ordered from lower to higher energy)
\begin{eqnarray}
\label{eq:E}
&&E(2M)=E(2M+2)=E_c\nonumber\\
&&E(2M+1)=\Delta\nonumber\\
&&E(2M-1)=E(2M+3)=4 E_c+\Delta\nonumber\\
&&E(2M+4)=E(2M-2)=9 E_c\nonumber\\
&&...,
\end{eqnarray}
where the ordering of the states changes if $E_c<\Delta/5$.
The fact that $E(2M+1)<(E(2M-1),E(2M+3))$ will be important for establishing CKE in Section~\ref{sec:kondo}.  

Since the ground state of the island is doubly degenerate, it behaves as a spin 1/2 system at low enough temperatures ($T<T^*$).
For later purposes we introduce a pseudospin label:
\begin{equation}
|2 M\rangle \equiv |\Downarrow_0\rangle \mbox{   ;   } |2M+2\rangle \equiv |\Uparrow_0\rangle,
\end{equation}
where the subscript $0$ simply labels the location of the island. 
From now on pseudospin and real spin degrees of freedom will be denoted by double and single arrows, respectively. 
This qubit is well-separated from a dense forest of many-body states, which have excitation energies greater than $\Delta-E_c$.
Due to entropic issues\cite{tinkham1996} that arise from having an approximately manifold degenerate set of excited states, 
\begin{equation}
\label{eq:T_star}
T^*\simeq \frac{\Delta-E_c}{\ln\left(\frac{\Delta-E_c}{\delta}\right)}
\end{equation}
is smaller than the excitation energy gap $\Delta-E_c$.

The mapping to a two-level system holds even when $M$ is macroscopically large, and has been amply corroborated by experiments.\cite{lafarge1993,eiles1993,hergenrother1994,nakamura1999}
The degeneracy between $|2M\rangle$ and $|2M+2\rangle$ may be lifted by tuning the gate voltage away from the degeneracy point, which is akin to applying a pseudospin magnetic field along $\hat z$, or by coupling the island to a superconducting reservoir, which acts as a pseudo-magnetic field in the $xy$ plane. 
The charge-Kondo effect discussed below emerges when these pseudospin magnetic fields are very weak. 

For completeness, we note that there is a twofold charge-degeneracy point even when $\Delta<E_c$.
In this case the degeneracy occurs between states that differ by a single electron,\cite{houzet2005} and therefore this problem
is adiabatically connected to that of Ref.~[\onlinecite{matveev1991}].

\section{Coupling to the Leads: Kondo Model}
\label{sec:kondo}

In this section we consider the influence of conducting leads on the superconducting island with charge degeneracy, at temperature $T<T^*$. 
We assume all non-thermal perturbations to be small compared to $\Delta-E_c$
Then the NSN heterostructure behaves as a localized pseudospin 1/2 interacting with a bath of itinerant fermions, and Eq.~(\ref{eq:model}) can be truncated on symmetry grounds into an effective Hamiltonian 
\begin{eqnarray}
\label{eq:h_eff_ps}
{\cal H}_{\rm eff}&=& {\cal H}_{\rm l}+ \sum_{\alpha\alpha'}\lambda_{0}^{\alpha\alpha'}S^0 \tilde{s}^0_{\alpha\alpha'}\nonumber\\
&+& \sum_{\alpha\alpha'} \left[\lambda_{||}^{\alpha\alpha'}S^z \tilde{s}^z_{\alpha\alpha'}+\lambda_{\perp}^{\alpha\alpha'}\left(S^+ \tilde{s}^-_{\alpha\alpha'}+{\rm h.c.}\right)\right],
\end{eqnarray}
where $2 S^0 =|\Uparrow_0\rangle\langle\Uparrow_0|+|\Downarrow_0\rangle\langle\Downarrow_0|$, $2 S^z=|\Uparrow_0\rangle\langle\Uparrow_0|-|\Downarrow_0\rangle\langle\Downarrow_0|$ and $S^+ =|\Uparrow_0\rangle\langle\Downarrow_0|$ are the pseudospin operators characterizing the state of the superconducting island, and
\begin{equation}
\tilde{s}^{i}_{\alpha\alpha'} = \sum_{k k' \sigma \sigma'} \tilde{c}^\dagger_{\alpha k \sigma} \tau^i_{\sigma \sigma'} \tilde{c}_{\alpha' k' \sigma'}
\end{equation}
denotes the pseudospin density in the leads ($i\in\{0,x,y,z\}$). 
In addition, ${\boldsymbol\tau}$ is a vector of Pauli matrices ($\tau^0$ is the identity matrix) and $\tilde{c}_{\alpha k \sigma}$ is a pseudospin operator defined as
\begin{equation}
\tilde{c}_{\alpha k \uparrow} = c_{\alpha k \uparrow}\mbox{   ;   } \tilde{c}_{\alpha k \downarrow} = c^\dagger_{\alpha,-k \downarrow}.
\end{equation}
Physically, $\tilde{s}^0$ is the spin density along $\hat{z}$, $\tilde{s}^z$ is the charge density and $\tilde{s}^\pm$ is the pairing operator. 
In Eq.~(\ref{eq:h_eff_ps}), the term involving  $S^0$ ($S^z$) describes normal spin-dependent (spin-independent) scattering and the term $\propto S^{\pm}$ describes Andreev processes whereby the number of Cooper pairs in the island changes by one; the {\em xy} symmetry is simply a reflection of gauge invariance. 
Ignoring real-spin-flips (which are unimportant at low energies), Eq.~(\ref{eq:h_eff_ps}) is the most general form of exchange interaction between the localized and itinerant pseudospins.
We have omitted perturbations that break charge degeneracy; these can be easily incorporated in Eq.~(\ref{eq:h_eff_ps}) as pseudospin Zeeman fields.

\begin{table*}[t]
\caption{Virtual elastic processes to second order in single-particle tunneling (the extension to fourth order processes is deferred to the Appendix). The initial and final states are in the truncated, low-energy Hilbert space whereas the intermediate states trespass into the high-energy sector. 
We ignore inelastic processes because they are suppressed at $T\ll T^*$.  
The energy of all initial and final states is $\simeq E_c$ because the most important tunneling events involve electrons/holes at the Fermi surface ($\xi_{k}\simeq\xi_{k'}\simeq 0$).}
% For a particle-hole symmetric superconducting energy spectrum, the processes in which the initial state has a hole in one of the leads are redundant and are omitted from this Table.}
\vspace{0.1 in}
\begin{tabular}{c c c c c}
\hline\hline
Initial State ($|i\rangle$) & Intermediate States ($|n\rangle$) & Final State ($|f\rangle$) & $E_i-E_n$ & $\langle f|{\cal H}_T|n\rangle\langle n|{\cal H}_T|i\rangle$\\
\hline\hline
$c^\dagger_{\alpha k\uparrow}|\o\rangle|2M+2\rangle$\mbox{    } & $|\o\rangle\gamma^\dagger_{n\uparrow} |2M+2\rangle$  & $c^\dagger_{\alpha' k'\uparrow}|\o\rangle|2M+2\rangle$ & $E_c-(4 E_c+\sqrt{\epsilon_n^2+\Delta^2})$ & $t^*_{\alpha k n} t_{\alpha' k' n} u_n^2 $ \\[1ex]
 & $c^\dagger_{\alpha' k' \uparrow} c^\dagger_{\alpha k \uparrow}|\o\rangle\gamma^\dagger_{n\downarrow}|2M+2\rangle$  &  & $E_c-\sqrt{\epsilon_n^2+\Delta^2}$ & $-t^*_{\alpha k n} t_{\alpha' k' n} v_n^2 $ \\[1ex]
\hline
$c^\dagger_{\alpha k \uparrow}|\o\rangle|2M\rangle$\mbox{    } & $|\o\rangle\gamma^\dagger_{n\uparrow} |2M\rangle$  & $c^\dagger_{\alpha' k'\uparrow}|\o\rangle|2M\rangle$ & $E_c-\sqrt{\epsilon_n^2+\Delta^2}$ & $t^*_{\alpha k n} t_{\alpha' k' n} u_n^2 $ \\[1ex]
 & $c^\dagger_{\alpha' k' \uparrow} c^\dagger_{\alpha k \uparrow}|\o\rangle\gamma^\dagger_{n\downarrow} |2M\rangle$  &  & $E_c-(4 E_c+\sqrt{\epsilon_n^2+\Delta^2})$ & $-t^*_{\alpha k n} t_{\alpha' k 'n} v_n^2 $ \\[1ex]
\hline
$c^\dagger_{\alpha k \uparrow}|\o\rangle|2M\rangle$ & $|\o\rangle\gamma^\dagger_{n\uparrow} |2M\rangle$ & $c_{\alpha' k' \downarrow}|\o\rangle|2M+2\rangle$ & $E_c-\sqrt{\epsilon_n^2+\Delta^2}$ & $-t^*_{\alpha k n} t^*_{\alpha' k' n} u_n v_n$    \\[1ex]
& $c^\dagger_{\alpha k \uparrow} c_{\alpha' k'\downarrow}|\o\rangle\gamma^\dagger_{n\downarrow} |2M+2\rangle$ & & $E_c-\sqrt{\epsilon_n^2+\Delta^2}$ & $-t^*_{\alpha k n} t^*_{\alpha' k' n} u_n v_n$\\[1ex]
\hline\hline
\label{tab:k2}
\end{tabular}
\end{table*}

$\lambda_{||,\perp}^{\alpha\alpha'}$ are the Kondo couplings.
Their microscopic expressions can be extracted perturbatively from Eq.~(\ref{eq:model}) via\cite{fiete2002} 
\begin{equation}
\langle f|{\cal H}_{\rm eff}|i\rangle\simeq \sum_{n} \frac{\langle f|{\cal H}_T|n\rangle\langle n|{\cal H}_T|i\rangle}{E_i-E_n},
\end{equation}
where $|i\rangle$ ($|f\rangle$) is the initial (final) state with energy $E_i$ ($E_f$) and $|n\rangle$ is an intermediate state with energy $E_n$ (see Table \ref{tab:k2}).
It follows that
\begin{eqnarray}
\label{eq:lambda}
\lambda^{\alpha\alpha'}_{0} &=& \sum_n \frac{\langle\Uparrow_{\alpha k};\Uparrow_0|{\cal H}_T|n\rangle\langle n|{\cal H}_T|\Uparrow_{\alpha' k'};\Uparrow_0\rangle}{E_{\Uparrow_{\alpha k};\Uparrow_0}-E_n}\nonumber\\
&+&\sum_n\frac{\langle\Uparrow_{\alpha k};\Downarrow_0|{\cal H}_T|n\rangle\langle n|{\cal H}_T|\Uparrow_{\alpha' k'};\Downarrow_0\rangle}{E_{\Uparrow_{\alpha k};\Downarrow_0}-E_n}\nonumber\\ 
\lambda^{\alpha\alpha'}_{||} &=& \sum_n \frac{\langle\Uparrow_{\alpha k};\Uparrow_0|{\cal H}_T|n\rangle\langle n|{\cal H}_T|\Uparrow_{\alpha'k'};\Uparrow_0\rangle}{E_{\Uparrow_{\alpha k};\Uparrow_0}-E_n}\nonumber\\
&-&\sum_n\frac{\langle\Uparrow_{\alpha k};\Downarrow_0|{\cal H}_T|n\rangle\langle n|{\cal H}_T|\Uparrow_{\alpha' k'};\Downarrow_0\rangle}{E_{\Uparrow_{\alpha k};\Downarrow_0}-E_n}\nonumber\\ 
\lambda^{\alpha\alpha'}_{\perp} &=& \sum_n \frac{\langle\Uparrow_{\alpha k};\Downarrow_0|{\cal H}_T|n\rangle\langle n|{\cal H}_T|\Downarrow_{\alpha' k'};\Uparrow_0\rangle}{E_{\Uparrow_{\alpha k};\Downarrow_0}-E_n},
\end{eqnarray}
 where $|\Uparrow_{\alpha k}\rangle\equiv c^\dagger_{\alpha k\uparrow}|\o\rangle$ and $|\Downarrow_{\alpha k}\rangle\equiv c_{\alpha k\downarrow}|\o\rangle$ represent spin-up electrons (spin-down holes) at the Fermi surface ($|\o\rangle$ stands for the filled Fermi sea in the leads). 
Table \ref{tab:k2} collects all the intermediate states $|n\rangle$, along with their energies $E_n$.

We proceed with the evaluation of the tunneling matrix elements using the BCS formalism.
For instance, the matrix elements for an electron or a hole tunneling from lead $\alpha$ to the $n$-th level in the island are
\begin{eqnarray}
\label{eq:tme}
\langle\Uparrow_{\alpha k};\Uparrow_0|{\cal H}_T \gamma^\dagger_{n\uparrow} |\Uparrow_0\rangle &\simeq& \langle\Uparrow_{\alpha k};\Downarrow_0|{\cal H}_T \gamma^\dagger_{n\uparrow}|\Downarrow_0\rangle \simeq t_{\alpha k n} u_n \nonumber\\
\langle\Downarrow_{\alpha k};\Uparrow_0|{\cal H}_T \gamma^\dagger_{n\uparrow} |\Uparrow_0\rangle &\simeq& \langle\Downarrow_{\alpha k};\Downarrow_0|{\cal H}_T \gamma^\dagger_{n\uparrow} |\Downarrow_0\rangle \simeq  -t^*_{\alpha k n} v_n\nonumber,
\end{eqnarray}
 where $\gamma^\dagger_{n\sigma}=u_n d^\dagger_{n\sigma}+\sigma v_n d_{n,-\sigma}$ is the BCS quasiparticle creation operator, $u_n^2=(1+\epsilon_n/\sqrt{\epsilon_n^2+\Delta^2})/2$ and $v_n^2=1-u_n^2$ are the standard BCS coherence factors (we assume $u_n$ and $v_n$ to be real without loss of generality because we consider only one superconductor).
Other matrix elements may be computed similarly.
Even though intermediate states like $\gamma^\dagger_{n\sigma}|2M+2\rangle$ do not have a well-defined number of particles, they are excellent approximations to the true eigenstates of the superconducting island when $\Delta>>\delta$.

The explicit expressions for the couplings can now be read out from Table \ref{tab:k2}:
\begin{eqnarray}
\label{eq:lambda_explicit}
&&\lambda_{||}^{\alpha\alpha'}=\sum_n \left(\frac{t^*_{\alpha n k} t_{\alpha' k' n}}{-3 E_c-\sqrt{\epsilon_n^2+\Delta^2}}-\frac{t^*_{\alpha n k} t_{\alpha' k' n}}{E_c-\sqrt{\epsilon_n^2+\Delta^2}}\right)\nonumber\\
&&\lambda_{\perp}^{\alpha\alpha'}=-2\sum_n t^*_{\alpha n k} t^*_{\alpha' k' n}\frac{u_n v_n}{E_c-\sqrt{\epsilon_n^2+\Delta^2}}.
\end{eqnarray}
It is easy to verify that $\lambda_{0}=0$ by invoking particle-hole symmetry for the energy spectrum of the superconducting island.
Since $\lambda_{||}$ is associated with normal scattering, it involves virtual transitions that do not flip pseudospins.
%In fact $\lambda_{||}$ is defined as the energy difference between a configuration in which the lead and island have parallel pseudospins and a configuration in which the two pseudospins are antiparallel.
The sign of $\lambda_{||}$ will be important in the considerations below.
$\lambda_{\perp}$ is associated to Andreev scattering, hence it involves virtual transitions that flip the pseudospins of both the island and the leads.
%In this case the parallel pseudospin configuration does not result in any contribution because $\langle\Uparrow_0;\Uparrow_{\alpha k}|S^+ \tilde{s}^-|\Uparrow_0;\Uparrow_{\alpha k}\rangle=\langle\Uparrow_0;\Uparrow_{\alpha k}|S^+ \tilde{s}^-|\Downarrow_0;\Downarrow_{\alpha k}\rangle=0$.
The sign of $\lambda_{\perp}$ is unimportant as it may be reversed by a gauge transformation.
%; our sign convention is set via $u_n v_n=\langle 2 M| d_{n\uparrow} d_{n\downarrow}|2 M+2\rangle$.
 
In principle, evaluating the sums in Eq.~(\ref{eq:lambda_explicit}) requires a detailed knowledge of the tunneling amplitudes $t_{\alpha n k}$, which in turn depend on details of the wavefunctions at the NS interfaces.
In practice, simplifying assumptions may be quantitatively satisfactory. 
For instance, $t_{\alpha n k}\simeq t_{\alpha n}$ is a good approximation for NS junctions that are atomically thin (recall that $k$ is the momentum perpendicular to the NS interface).
This renders $\lambda^{\alpha\alpha'}_{||,\perp}$ independent of $k$ and $k'$.
Similarly, we conceive $t^*_{\alpha n} t_{\alpha' n} \simeq t^*_\alpha t_{\alpha'}$ and $t^*_{\alpha n} t^*_{\alpha' n} \simeq t^*_\alpha t^*_{\alpha'}$  with the understanding that the $n-$dependence is weak for states that are contained within a very narrow strip around the Fermi energy.
This approximation is motivated by the fact that the main contributions to $\lambda_{||}$ and $\lambda_\perp$ in Eq.~(\ref{eq:lambda_explicit}) originate from states with $|\epsilon_n|\lesssim E_c<E_T$ and $|\epsilon_n|\lesssim \Delta<E_T$, respectively. 
%The latter assumption may be more delicate than the former, because of randomness in the single-particle states of the superconducting island. Granted, a statistical average over different realizations of the random potential in the superconductor yields a zero mean for the tunneling amplitudes $t_{\alpha n}$.\cite{aleiner2002,glazman2005} 
%However, it seems reasonable that the statistical values of the products $t_{\alpha n} t_{\alpha' n}$ and $t_{\alpha n} t^*_{\alpha' n}$ will be largely independent of $n$ if we limit ourselves to states that are very close to the Fermi energy.

\begin{figure}[h]
\begin{center}
\includegraphics[scale=0.3,angle=270]{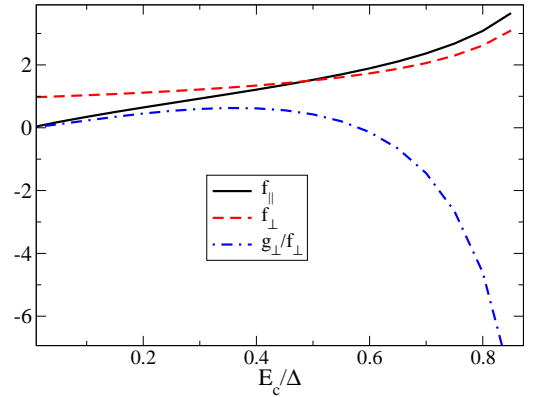}
\caption{Parameters $f_\perp$, $f_{||}$ and $g_\perp$ as a function of $E_c/\Delta$.
%, in units of the small dimensionless parameter $\Gamma/\delta$. 
 These parameters (defined in the main text and in the Appendix) determine the values of the Kondo couplings at an energy scale $T^*$ (see Eqs.~(\ref{eq:T_star}) and ~(\ref{eq:l_T0})). 
When $E_c$ is sufficiently close to $\Delta$ (precisely how close is contingent on the value of $\Gamma/\delta$)  the bare couplings become large, which invalidates the perturbative renormalization group approach employed in this work. 
%Moreover, as $E_c\to\Delta$ the pseudospin 1/2 approximation is valid only if $T\to 0$. 
For typical mesoscopic superconducting islands $E_c/\Delta\simeq O(1)$ and therefore $\lambda_{||}(T^*)/\lambda_{\perp}(T^*)\simeq O(1)$.} 
\label{fig:fg}
\end{center}
\end{figure}

Under the above assumptions, the tunneling amplitudes may be taken to be real and the Kondo couplings read
\begin{equation}
\label{eq:lambda_coh}
\lambda_{||,\perp}^{\alpha\alpha'}=\lambda_{||,\perp}^{LL}\left(\begin{array}{cc} 1 & t_R/t_L\\t_R/t_L & (t_R/t_L)^2\end{array}\right).
\end{equation}
Eq.~(\ref{eq:lambda_coh}) may be diagonalized with a unitary transformation,\cite{glazman1988} which converts Eq.~(\ref{eq:h_eff_ps}) into a single-channel Kondo model:
\begin{equation}
\label{eq:sckm}
{\cal H}_{\rm SCK}={\cal H}_{\rm l}+ \left[\lambda_{||} S^z \tilde{s}^z_1+\lambda_{\perp}\left(S^+ \tilde{s}^-_1+{\rm h.c.}\right)\right],
\end{equation} 
where $\tilde{s}^{i}_1 =\sum_{k,k'} \tilde{c}^\dagger_{1 k \sigma} \tau^i_{\sigma \sigma'}\tilde{c}_{1 k' \sigma'}$ and 
\begin{eqnarray}
\label{eq:Psi}
\tilde{c}_{1 k\sigma}&=&(t_L \tilde{c}_{L k\sigma}+t_R \tilde{c}_{R k \sigma})/\sqrt{t_L^2+t_R^2}\nonumber\\
%\left(\begin{array}{c} \tilde{c}_{1 k \sigma} \\ \tilde{c}_{2 k \sigma}\end{array}\right)=\frac{t_R}{\sqrt{t_R^2+t_L^2}}\left(\begin{array}{cc} \frac{t_L}{t_R} & 1 \\ 1 & -\frac{t_L}{t_R}\end{array}\right)\left(\begin{array}{c}\tilde{c}_{L k \sigma} \\ \tilde{c}_{R k \sigma}\end{array}\right)\nonumber\\
\lambda_{||,\perp} &=& \lambda_{||,\perp}^{LL}+\lambda_{||,\perp}^{RR}% \mbox{   ;   } \lambda_{||,\perp}^{(2)}= 0.
\end{eqnarray}
Thus only one out of the two conduction channels is coupled to the superconductor.
Starting from  Eq.~(\ref{eq:tme}) and using $\epsilon_n=n \delta$ we derive explicit expressions for the Kondo couplings at second order in tunneling amplitude: 
\begin{equation}
\label{eq:lambda_2_0}
\nu \lambda_{\perp} =\frac{\Gamma}{\delta} f_\perp(x) \mbox{   ;    }\,\,
\nu \lambda_{||} =\frac{\Gamma}{\delta} f_{||}(x),
\end{equation}
where $x\equiv E_c/\Delta$,
\begin{eqnarray}
\label{eq:lambda_2}
f_{\perp} (x) &\simeq& \frac{1}{\pi} \frac{4}{\sqrt{1-x^2}}\tan^{-1}\sqrt{\frac{1+x}{1-x}}\nonumber\\
f_{||} (x) &\simeq& x f_{\perp}+\frac{1}{\pi} \frac{12 x}{\sqrt{1-9 x^2}}\tan^{-1}\sqrt{\frac{1-3 x}{1+3 x}},
\end{eqnarray}
 $\nu$ is the density of states of the leads at the Fermi energy and $\Gamma=\pi\nu(t_L^2+t_R^2)$ denotes the broadening of the energy levels in the dot due to its coupling to the leads.
$\Gamma$ is related to a dimensionless parameter representing the transparency of the tunnel junctions:
\begin{equation}
\label{eq:g0}
g_T = g_L+g_R = \frac{(t_L^2+t_R^2)\nu}{\delta} =\frac{\Gamma}{\pi \delta}\ll 1,
\end{equation}
where $g_L$ and $g_R$ are the conductances of the left and right tunnel junctions in units of $2\pi e^2/\hbar$.
%Due to the coherent propagation of electrons in the dot, the characteristic conductance $g$ corresponds to that of two resistors connected in parallel rather than in series, and $1/\delta$ is the single-particle density of states of the island at the Fermi level.
In the derivation of Eq.~(\ref{eq:lambda_2}) we approximated $\sum_n$ by $\int_{-\infty}^{\infty} dn$; this is justified in mesoscopic superconducting islands, where $\delta\lesssim 10^{-3}\Delta$.
In Fig.~(\ref{fig:fg}) we plot $f_{\perp}$ and $f_{||}$.
An essentially identical expression for $\lambda_\perp$ was found by Hekking {\em et al.} in Ref.~[\onlinecite{hekking1993}];
however, these authors did not contemplate the possibility of a Kondo effect and did not evaluate $\lambda_{||}$, whose positive sign is important in order to produce CKE.

As demonstrated in the Appendix, Kondo correlations become apparent only when we carry out perturbation theory to {\em fourth} order in tunneling.
Indeed at fourth order both coupling constants $\lambda_\perp$ and $\lambda_{||}$ develop infrared singularities, in a manner that is consistent with the renormalization group equations of the anisotropic Kondo model:\cite{hewson1993,coleman} 
\begin{eqnarray}
\label{eq:kt}
\frac{d\lambda_{||}}{d l}&=&2 \nu \lambda_{\perp}^2-2\nu^2\lambda_{||}\lambda_{\perp}^2\nonumber\\
\frac{d\lambda_{\perp}}{d l}&=& 2 \nu \lambda_\perp \lambda_{||}-\nu^2\lambda_{\perp}(\lambda_{||}^2+\lambda_{\perp}^2),
\end{eqnarray}
where $l\equiv\ln(T^*/T)$ and by definition $\lambda_{||,\perp}(l=0)\equiv\lambda_{||,\perp}(T^*)$ are the values of the Kondo couplings at an energy scale $T^*$, below which the Kondo model becomes applicable. 
The present analysis is quantitatively reliable provided that $\lambda_{||,\perp}(T^*)\ll 1$.
%These initial values are not given by Eq.~(\ref{eq:lambda_2}); instead, 
We show in the Appendix that
\begin{eqnarray}
\label{eq:l_T0}
\nu\lambda_\perp (T^*) &\simeq& \frac{\Gamma}{\delta} f_{\perp}\left(1+\frac{\Gamma}{\delta}\frac{g_\perp}{f_\perp}\right)+ 2\frac{\Gamma^2}{\delta^2} f_{||} f_\perp\ln\frac{\Delta}{T^*}\nonumber\\
\nu\lambda_{||}(T^*) &\simeq& \frac{\Gamma}{\delta} f_{||} \left(1+\frac{\Gamma}{\delta} \frac{g_{||}}{f_{||}}\right)+2\frac{\Gamma^2}{\delta^2} f_\perp^2\ln\frac{\Delta}{T^*},
\end{eqnarray}
where $g_\perp$ and $g_{||}$ are dimensionles functions of $\Delta/E_c$ (the former is plotted in Fig.~(\ref{fig:fg})).
% and $g_{||}$ is another dimensionless function of $\Delta/E_c$ which we have not evaluated.
Eq.~(\ref{eq:l_T0}) is valid provided that the $O(\Gamma^2)$ terms are smaller than the $O(\Gamma)$ terms; for $\Gamma/\delta\to 0$ Eq.~(\ref{eq:l_T0}) reduces to Eq.~(\ref{eq:lambda_2_0}). 
%$(\Gamma/\delta) g_\perp/f_\perp\ll 1$ and $(\Gamma/\delta) g_{||}/f_{||}\ll 1$; if these conditions are not met, higher order tunneling terms start to become important.
The logarithmic terms in Eq.~(\ref{eq:l_T0}) suggest that the Kondo couplings begin to renormalize from energy scales of order $\Delta$.
%{\bf did we consider just one lead for the 4th order calculation?} 
$\lambda_{||}(T^*)>0$ ensures that both $\lambda_{||}$ and $\lambda_{\perp}$ will flow to strong coupling through Eq.~(\ref{eq:kt}).
The energy scale at which ${\rm max}\{\nu\lambda_\perp(l_c),\nu\lambda_{||}(l_c)\}\simeq 1$ is given by the Kondo temperature: 
\begin{equation}
\label{eq:TK}
T_K=T^*\exp(-l_c),
\end{equation}
where $T^*$ should be replaced by $8 E_c$ when $E_c<T^*/8$ (recall Eq.~(\ref{eq:E})).
%We tailor our criterion so that when $\lambda_{\perp}(0)=\lambda_{||}(0)\equiv\lambda$ we obtain $1/l_c\simeq 2\nu \lambda+\log(\sqrt{2\lambda})$; the latter yields the textbook\cite{coleman} expression for $T_K$ in the isotropic Kondo case.  
%Quantitative estimates of the Kondo temperature in NSN devices are relegated to Section~\ref{sec:disc}.

\section{Discussion}
\label{sec:disc}
%{\bf I'm concerned that we'll never have a single channel case except if we use a point contact}

In this section we derive quantitative estimates for the charge-Kondo temperature in NSN heterostructures,  examine the influence of multi-channel junctions, propose various experimental NSN platforms wherein CKE might be observed, and compare our proposal of CKE in NSN heterostructures with previous proposals of CKE in other systems.

\subsection{Estimates for the Charge-Kondo Temperature}

The Kondo temperature of our system can be quantified combining Eqs.~(\ref{eq:lambda_2_0}),~(\ref{eq:lambda_2}),~(\ref{eq:kt}),~(\ref{eq:l_T0}) and ~(\ref{eq:TK}).
For pedagogical purposes we begin by discussing a few particular cases that can be solved analytically.

\begin{figure}[h]
\begin{center}
\includegraphics[scale=0.3, angle=270]{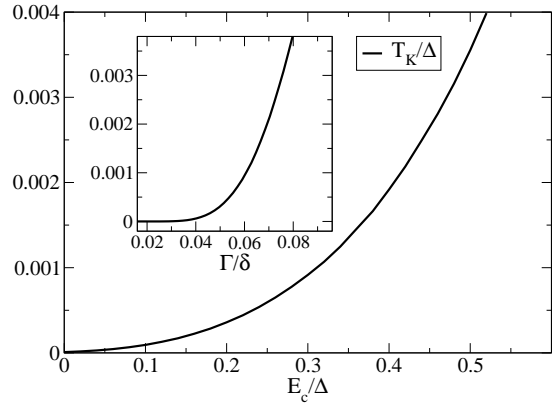}
\caption{Kondo temperature (Eq.~(\ref{eq:TK})) (in units of $\Delta$) as a function of $E_c/\Delta$, for $\Gamma/\delta\simeq 0.08$. We choose $\Delta/\delta\simeq 1000$, although the results are only weakly sensitive to $\Delta/\delta$. For this set of parameters, the Kondo temperature can reach $\sim 10 {\rm mK}$ by selecting $E_c/\Delta$ in the appropriate range. The decrease in $T^*$ as $E_c/\Delta$ increases is overcompensated by the gain in the bare Kondo coupling. 
In our estimates we have ignored both $g_\perp$ and $g_{||}$.
The perturbative analysis breaks down when $\nu \lambda_{||,\perp}(T^*)\sim O(1)$.
%${\rm max}\{(\Gamma/\delta)f_{||}\ln(\Delta/T^*),(\Gamma/\delta)(f_\perp^2/f_{||})\ln(\Delta/T^*)\}\sim O(1)$.   
For $\Gamma/\delta\simeq 0.08$ this breakdown is apparent at $E_c/\Delta\gtrsim 0.5$.
{\em Inset:} Kondo temperature (in units of $\Delta$) as a function of $\Gamma/\delta$, for $E_c=0.5\Delta$.}
\label{fig:TK}
\end{center}
\end{figure}

The first special case corresponds to $E_c\simeq 0.5 \Delta$, for which $\lambda_\perp(T^*)\simeq\lambda_{||}(T^*)$ and 
\begin{eqnarray}
\label{eq:tk_1}
T_K(\Delta\to 2 E_c) &\simeq& T^*\exp(-1/2\nu\lambda_\perp)\nonumber\\
&\simeq&\Delta\exp(g_\perp)\exp\left(-\frac{\delta/\Gamma}{2 f_\perp}\right), 
\end{eqnarray}
where for simplicity we have neglected third order terms in the RG equations.
In the second line of Eq.~(\ref{eq:tk_1}) we have made a Taylor expansion under the assumption that $(\Gamma/\delta) g_\perp/f_\perp\ll 1$ and $2\nu\lambda_\perp\ln(\Delta/T^*)\ll 1$.
For mesoscopic superconductors, $\ln(\Delta/T^*)\lesssim 2$ and hence the latter condition requires that $\Gamma/\delta<0.1$.
A typical Kondo temperature in this case (using $f_\perp\simeq 1.4$) reads $T_K\sim 10^{-3}\Delta\sim 1-10 {\rm mK}$ for conventional superconductors.
It is interesting that even though the Kondo Hamiltonian is valid at energy scales below $T^*$, the prefactor of the Kondo temperature in Eq.~(\ref{eq:tk_1}) is effectively shifted from $T^*$ to $\Delta$.
An analog situation occurs between $E_c$ and $\delta$ in SKE.\cite{garate2011}  

Another simple limit is that of $E_c\ll\Delta$, in which case $\lambda_{||}(T^*)\ll\lambda_\perp(T^*)$ and accordingly\cite{gogolin}
\begin{equation}
\label{eq:tk_2}
T_K (\Delta\gg E_c)\simeq T^*\exp(-\pi/4\nu\lambda_\perp),
\end{equation}
where once again we have neglected third order terms in Eq.~(\ref{eq:kt}). 
The exponent of Eq.~(\ref{eq:tk_2}) is a factor $\pi/2$ more negative than the exponent of Eq.~(\ref{eq:tk_1}), and $f_\perp(E_c\ll\Delta)\simeq 1<f_\perp(E_c\simeq 0.5\Delta)$.
Consequently, for a given value of $\Gamma/\delta$, $T_K (\Delta\gg E_c) \ll T_K(\Delta\simeq 2 E_c)$.
% for the highly anisotropic case (Eq.~(\ref{eq:tk_2})) is much lower than for the isotropic case (Eq.~(\ref{eq:tk_1})).

A third special case consists of $\lambda_{||}(T^*)>>\lambda_\perp(T^*)$.
 However, in our model this regime does not arise for any value of $E_c/\Delta$ and thus we disregard it.

We now solve more general cases numerically.
 Fig.~(\ref{fig:TK}) displays the charge-Kondo temperature for a generic range of parameters.
For fixed $\Gamma/\delta$, $T_K$ increases rapidly with $E_c/\Delta$ (at least up until $E_c/\Delta\simeq 0.6$).
%We find $T_K\simeq 0.003 \Delta$ for $\Gamma/\delta\simeq 0.08$ and $E_c/\Delta\simeq 0.5$.
Our findings establish $T_K\simeq 10 {\rm mK}$ for superconductors such as Al or Ta, and $T_K\simeq 20 {\rm mK}$ for superconductors such as Nb and Pb.
%Parametrically larger values of the Kondo temperature might be obtained in unconventional superconductors with fully gapped Fermi surfaces, such MgB$_2$ and (possibly) iron pnictides. {\bf BUT THESE HAVE TINY COHERENCE LENGTHS}

As $E_c/\Delta\to 1$, higher order tunneling processes (encoded in $g_{||,\perp}$) become important, $\nu\lambda_{\perp,||}(T^*)$ is no longer small and correspondingly our perturbative calculation of $T_K$ becomes unreliable.
%Of course, the fact that perturbative approaches break down does not mean that there is no Kondo effect.
%In fact, in the strong coupling regime one expects the Kondo temperature to be a sizeable fraction of the ultraviolet cutoff $T^*$, 
The divergence of $g_\perp$ towards {\em negative} values (Fig.~\ref{fig:fg}) casts some doubt on whether the Kondo temperature will continue to increase as $E_c/\Delta\to 1$. 

In the inset of Fig.~(\ref{fig:TK}) we illustrate the dependence of $T_K$ on $\Gamma/\delta$.
For a given value of $E_c/\Delta$, $T_K$ increases exponentially with $\Gamma/\delta$.
%We find $T_K\simeq 10^{-3}\Delta$ for $E_c\simeq 0.5 \Delta$ and $\Gamma/\delta\simeq 0.06$.
Only small values of $\Gamma/\delta$ are allowed in our calculation, which treats tunneling as a weak perturbation.
In the strong tunneling regime ($\Gamma\gg\delta$) the charge of the superconducting island fluctuates strongly and it is {\em a priori} unclear what the fate of CKE will be. 
This issue will be addressed in Section~\ref{sec:bozo}.

\subsection{Influence of Multichannel Leads}

Thus far we have been considering only one conduction channel per lead.
This is appropriate for atomic point contacts, wherein itinerant electrons are fully characterized by their momentum in the direction perpendicular to the NS interface. 
Nonetheless, due to the small Fermi wavelength of metals, most NS contacts contain multiple channels that describe e.g. transverse momenta of electrons in the lead.
Multichannel leads can be readily incorporated into Eq.~(\ref{eq:model}) by rewriting the tunneling Hamiltonian and the Hamiltonian for the leads:
\begin{eqnarray}
\label{eq:multichannel}
{\cal H}_{\rm l} &=& \sum_{k \sigma i} \xi_k c^\dagger_{k \sigma i} c_{k \sigma i}\nonumber\\
{\cal H}_T &=& \sum_{k \sigma n i} t_{k n i} c^\dagger_{k \sigma i} d_{n \sigma} +{\rm h.c.},
\end{eqnarray} 
where $i=1,...,2 N$ is a channel index that includes the which-lead label.
In wide junctions the tunneling amplitude depends on the channel index $i$.
In contrast, we have taken the kinetic energy of the itinerant fermions to be independent of $i$; this can always be ensured by proper rescaling of the itinerant fermion operators.
In addition, we have assumed that each energy level in the superconducting dot remains only twofold degenerate, due to the chaotic motion of electrons in the dot.

In principle it is possible that only one linear combination of the channels couple to the superconductor. 
This is the case when any one of the following assumptions (from more to less stringent) is satisfied at energies $|\epsilon_n|\lesssim\Delta$: 
(i) $t_{k n i}\simeq t_{k i}$ (energy-independent magnitude and sign of the tunneling amplitude), (ii) $t_{k n i}\simeq a_n b_{k i}$ (separable tunneling amplitude), (iii) $t_{k n i} t_{k' n j}\simeq t_{k i} t_{k j}$ (energy-independent magnitude of the tunneling amplitude, and coherent tunneling through different junctions due to $L<\zeta$). 
%for $|\epsilon_n|\lesssim\Delta$ or equivalently $t({\bf r},{\bf r}')=t'({\bf r}) t''({\bf r}')$, where ${\bf r}$ and ${\bf r'}$ are two dimensional position vectors defined at the NS interface.
With this proviso one can always make a unitary trasformation in the spirit of Eq.~(\ref{eq:Psi}) and recover an effective one-channel Kondo model.

When the assumptions (i)-(iii) above fail, a multichannel Kondo effect invariably follows.
In NNN devices $T_K$ drops exponentially\cite{zarand2000} when the number of channels in the normal leads {\em and} in the normal metallic grain is much larger than one.
This argument does not apply to NSN devices insofar as 
the superconducting island is describable by random matrix theory and $L\lesssim\zeta$.
%its linear dimensions are smaller than the superconducting coherence length.
However, even in our case $T_K$ is generically lower for the multichannel than for the single-channel Kondo model.
Suppose for instance that there are $N\gg 1$ separate Kondo couplings of similar magnitude. 
In order to maintain the Coulomb blockade, each bare Kondo coupling must scale like $\sim 1/N$, i.e. each Kondo coupling must be very small.
Eq.~(\ref{eq:kt}) then dictates that the Kondo temperature will be very low.
This state of affair changes whenever a unitary transformation can map multichannel leads onto an effective single-channel lead, because in that case the only nonzero Kondo coupling in the transformed basis scales as the {\em sum} of all the original couplings (see Eq.~(\ref{eq:Psi}) for a $N=2$ example of this). 

In order to arrive at a single-channel Kondo problem in presence of wide NS junctions it is helpful that the Kondo couplings be of second order in the tunneling amplitude, because condition (iii) above is more realistic than either (i) or (ii).
In case of CKE in non-superconducting dots\cite{zarand2000} the $xy$-component of the Kondo coupling is $\sim O(t)$, whereas the $z-$component is $\sim O(t^2)$. 
In such situation one is generally left with a multichannel Kondo problem. 
This is a potentially important difference between CKE in NSN and NNN heterostructures.

\subsection{Possible Experimental Setups}

\begin{figure}[h]
\begin{center}
\includegraphics[scale=0.3]{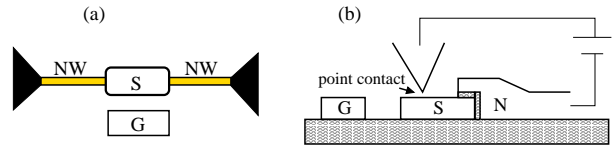}
\caption{Two possible arrangements for the experimental observation of the charge-Kondo effect in NSN systems. S=superconducting island, G=gate, N=normal metal, NW=semiconducting nanowire.
(a) Metallic-semiconducting hybrid setup (top view). The width of the semiconducting nanowires is comparable to their Fermi wavelength, which guarantees the emergence of a single-channel Kondo model at low energies.  Local gates placed at the nanowire-superconductor contacts can tune the channel transparency. The nanowires are long enough so that their energy-level spacing is negligible.
(b) All-metallic setup (side view). The atomic point contact fits a single conduction channel, whereas the wider NS junction may contain multiple channels. When $L\gg\zeta$, $T_K$ may be maximized by increasing the tunneling amplitude at the atomic point contact relative to that at the wide junction.} 
\label{fig:setup}
\end{center}
\end{figure}

Can the charge Kondo effect proposed in this paper be measured in experimentally realizable NSN devices? 
In a preceding subsection we have estimated charge Kondo temperatures of $\simeq 10-20 {\rm mK}$ for the single-channel Kondo model and reasonable values of $\Delta$, $\Gamma/\delta$ and $E_c$. 
This temperature is close to the lowest temperatures achieved in current dilution fridge refrigerators ($\lesssim 20 {\rm mK}$). 
Moreover, $T_K$ may in principle be made higher by designing more transparent junctions, tuning $\Delta/E_c$ closer to one or choosing materials with larger values of $\Delta$.
Thus while it will be difficult to experimentally access the unitary regime $T\ll T_K$, measurements conducted in the weak coupling regime $T\gtrsim T_K$ may be within reach.
%should already showcase fingerprints of Kondo correlations (these will be characterized in the next section).

Yet in spite of the extensive experimental work on mesoscopic NSN systems, there is no report of any CKE.
Some trivial reasons for this include (i) the effective temperature of the device being too high, (ii) the gate voltage not being tuned with sufficient precission to the charge degeneracy point, and (iii) purposely breaking the charge degeneracy by coupling the island to a superconducting reservoir.
A more fundamental reason might be that the NS junctions fabricated in typical experiments lead to multichannel Kondo problems with concomitantly lower $T_K$.

Next we discuss the experimental platform and requirements to observe CKE in NSN heterostructures.
First and foremost we need $\Delta>E_c$, which precludes ultrasmall superconducting islands due to their dominant energy.\cite{delft2001}
%Cooper pair boxes may provide viable examples of superconducting islands that constitute a pseudospin 1/2 quantum impurity, manipulable by gate voltages and Josephson couplings.\cite{nakamura1999}
%In Al-based boxes\cite{hergenrother1994,nakamura1999} $\Delta\simeq 0.2 {\rm meV}>E_c\simeq 0.1 {\rm meV}$.
Second, the superconducting island must be placed in the vicinity of a gate electrode. 
In order to measure CKE, the resolution of the gate voltage must be such that 
%\begin{equation}
$N_g-N_g^{(0)}= C_g (V_g-V_g^{(0)})/e\lesssim T_K/E_c$,
%\end{equation}
 where $N_g^{(0)}=2M+1$ is the charge degeneracy point (for any integer $M$), and $V_g^{(0)}=e N_g^{(0)}/C_g$ is the corresponding gate voltage.
For typical values of the gate capacitances ($C_g\simeq 10 {\rm aF}$) and the Kondo temperature ($T_K/E_c\simeq 10^{-3}$), it follows that the gate voltage resolution needs to be better than $10 \mu{\rm V}$.
This is achievable in current experimental devices.
If $N_g-N_g^{(0)}\gtrsim T_K/E_c$, the Kondo RG equations (Eq.~(\ref{eq:kt})) are cutoff at energy scales $\sim E_c (N_g-N_g^{(0)})\gtrsim T_K$ and the charge Kondo effect will not be fully developed.
However, even in this case there should be observable fingerprints of Kondo correlations (these are discussed in Section~\ref{sec:cond}).

Another desirable trait of the superconducting island is that it have no spatial symmetries and that its linear dimensions $L$ be smaller than the superconducting coherence length $\zeta$.
This ensures having only one conduction channel in the dot at any given energy, which increases the likelihood of producing a single-channel Kondo model at low energies.
Moreover, smaller island sizes mean a smaller ratio $\Delta/\delta$ and thus a larger value of $T^*$, which enhances the ultraviolet energy cutoff for the Kondo problem.
In principle, having a small-sized superconductor might conflict with the requirement that $\Delta>E_c$.
In practice,  there appears to be enough room in parameter space to ensure that $\Delta>E_c$ and $L<\zeta$ are satisfied simultaneously.
For instance, the best developed material is aluminum, whose BCS coherence length in the ballistic limit is $\zeta_0=\hbar v_F/(\pi\Delta)\simeq 1.5 \mu{\rm m}$.
For a mean free path of $l\simeq 15 {\rm nm}$, the actual coherence length is $\zeta=(\hbar D/\Delta)^{1/2}\simeq (\zeta_0 l)^{1/2}\simeq 150 {\rm nm}$. 
Thus an Al island of $L\simeq 80 {\rm nm}$ connected to a normal metal with junction area $A\simeq (80 {\rm nm})^2$ and $\epsilon\simeq 12$ (dielectric constant of AlO$_{\rm x}$) yields $E_c\simeq 0.12 {\rm meV}<\Delta\simeq 0.2 {\rm meV}$.
Niobium is another attractive material, due to its higher critical temperature.  
Although its small BCS coherence length ($\zeta_0\simeq 40 {\rm nm}$) requires relatively small and clean grains, 
the large dielectric constant of NbO$_{\rm x}$ ($\epsilon\simeq 40$) can produce sufficiently small charging energies even for $L\simeq 25  {\rm nm}$ and $A\simeq (25 {\rm nm})^2$ ($E_c\simeq 0.4 {\rm meV}<\Delta\simeq 1.4 {\rm meV}$).
Tantalum and lead can also be interesting candidates, as they have BCS gaps that are larger than that of Al and coherence lengths that are longer than that of Nb.
In Pb, $\zeta_0\simeq 90 {\rm nm}$ and $\zeta\simeq 35 {\rm nm}$ for $l\simeq 15 {\rm nm}$.
With $L\simeq 30 {\rm nm}$, $A\simeq (30 {\rm nm})^2$ and $\epsilon\simeq 25$ (dielectric constant of PbO$_{\rm x}$) we get $E_c\simeq 0.4 {\rm meV}<\Delta\simeq 1 {\rm meV}$.
Note that these figures are merely representative.
% and in general $A$ (the junction area) need not equal $L^2$ (see Fig.~\ref{fig:setup}). 

Last but not least, it is important that the NS interfaces have a small number of effective conduction channels.
We have previously discussed how a single-channel Kondo model may be plausible even in wide metallic junctions with $A\gg \lambda_F^2$, where $\lambda_F$ is the Fermi wavelength of the metal. 
In Fig.~\ref{fig:setup}a we sketch an alternative setup that combines a superconducting grain with semiconducting nanowires.
This setup is motivated by recent success in manufacturing semiconducting/superconducting nanocontacts.\cite{xiang2006,doh2008}
The large Fermi wavelength in semiconductors facilitates single-channel NS contacts without making the charging energy of the island larger than the superconducting gap.
In addition, the possibility of tuning Schottky barriers in semiconductor/metal interfaces by local gates creates prospects for measurable charge-Kondo temperatures.
 
Fig.~\ref{fig:setup}b constitutes yet another all-metallic experimental setup, which allows to reach sizeable $T_K$ in NSN devices with $L\gg\zeta$ (where electrons cannot coherently propagate from one junction to another).
One of the junctions is made from an atomic point contact, while the other junction is wide enough to ensure that the charging energy of the device remains less than the superconducting gap.
If the single channel at the point contact is much more strongly coupled to the superconductor than each of the multiple channels at the wide junction, then the charge-Kondo temperature is close to that of the single-channel Kondo model.
A theoretically straightforward way to achieve atomic point contacts is to use scanning tunneling microscopes (STM).
STM tips operated in the point contact mode can\cite{wei1998} capture Andreev tunneling (which is altered by charge-Kondo correlations), 
and recent STM experiments\cite{nishio2008} have succeeded measuring the density of states of ultrathin superconducting Pb grains.
An added challenge here is that the STM would have to operate at milliKelvin temperatures.
Alongside milliKelvin STMs, mechanically controllable single-channel quantum point contacts are being developed in superconducting systems.\cite{scheer1997}

\subsection{Comparison to Other Charge-Kondo Effects}

\begin{table*}[t]
\caption{Comparison between different charge-Kondo effects (CKE) and the spin-Kondo effect (SKE). 
%NS (NN) stands for a superconducting (normal) grain connected to a normal metallic lead, $NU$ stands for systems doped with negative-U molecules and 
QD stands for semiconducting quantum dots. 
$\omega$ is the largest of a variety of {\em extrinsic} energy scales such as temperature, bias voltage, pseudo-magnetic fields and the frequency of an external perturbation.}
%The re of the parameters are defined in the text.}  
%$T^*$ is the UV energy cutoff in NSN systems, $\delta$ is the single-particle energy-level spacing in the dot, $\Gamma\ll\delta$ is the broadening in the energy-levels of the dot due to hybridization with its surroundings.}
\vspace{0.1 in}
\begin{tabular}{|| c | c | c | c |c ||}
\hline\hline
& CKE in NSN& CKE in NU & CKE in NNN& SKE in QD \\
\hline\hline
Two-level system & $N$ or $N+1$ Cooper pairs & $N$ or $N+2$ electrons & $N$ or $N+1$ electrons & $\uparrow$ or $\downarrow$ electron\\[1ex]
\hline
Ultraviolet cutoff ($\Lambda$) & $T^*(\delta\ll T^*<E_c)$ & $|U|$ & $E_c (\gg\delta)$ & $\delta(\ll E_c)$\\[1ex]
\hline
Infrared cutoff  & $\omega$ & $\omega$ & $\max\{\omega,\delta\}$ & $\omega$\\[1ex]
\hline
$\nu\lambda_\perp(\Lambda)$ & $\sim (\Gamma/\delta) f_\perp$ &$\sim \Gamma/|U|$ & $\sim(\Gamma/\delta)^{1/2}$ & $\sim \Gamma/E_c$\\[1ex]
\hline
$\nu\lambda_{||}(\Lambda)$ & $\sim (\Gamma/\delta) f_{||}$ & $=\nu\lambda_\perp(\Lambda)$ & $0$ & $=\nu\lambda_\perp(\Lambda)$\\[1ex]
\hline
$T_K/\delta$ & unrestricted  & $\simeq 0$ & needs to be large & needs to be small\\[1ex]
\hline
Number of channels & $\geq 1$ & $1$ & $\geq 2 $ & $\geq 1$\\%[1ex]
\hline\hline
\end{tabular}
\label{tab:ckes}
\end{table*}

Part of our motivation for exploring NSN systems is to establish a ``quantum-dot counterpart'' for the charge-Kondo effect that is believed to arise in compounds doped with negative-U molecules (``NU'' systems). 
%As mentioned in the Introduction, there exists an alternative proposal for CKE in {\em non-superconducting} quantum dots that are coupled to metallic leads (NNN systems). 
In this section we discuss similarities and differences, as well as advantages and disadvantages of CKE in NU, NNN and NSN settings. 
%When useful, we make contact with the ordinary spin-Kondo effect in semiconducting quantum dots.
Some of the highlights are summarized in Table \ref{tab:ckes}.

The ultraviolet (UV) cutoff $\Lambda$ is the energy scale below which the Kondo model applies. 
It is lowest in NSN systems, where $T^*\ll \Delta$ due to entropic arguments.
This is partly why materials doped with negative-U centers reportedly show $T_K\simeq 10{\rm K}$, while we find $T_K\sim 10 {\rm mK}$ in mesoscopic NSN heterostructures.  
Note that CKE in NNN systems does not suffer from entropic issues because the ground state and the lowest excited state are both surrounded by similarly dense sets of single-particle states.

The infrared (IR) cutoff $\omega$ is the energy scale at which the renormalization of the Kondo couplings stops.
In NSN and NU systems $\omega$ is the largest amongst the temperature $T$, bias voltage $V$, and pseudo-magnetic fields $h_z,h_x$.
The full Kondo effect is observable only if $T_K>\omega$. 
In NNN systems, Kondo-like IR divergences originate from sums involving single-particle states {\em in the dot}.
Consequently $\delta$ too plays the role of an IR cutoff and the Kondo effect in NNN systems is observable only if $T_K>{\rm max}\{\delta,\omega\}$.
$T_K>\delta$ is difficult to satisfy in semiconducting dots.

Concerning the Kondo couplings, in NU systems they are isotropic in pseudospin space and originate from a single energy level.
In NSN and NNN systems, the Kondo couplings are anisotropic and sensitive to the multilevel energy spectrum of the dot.
While the anisotropy becomes gradually less pronounced at lower energies, it still plays an essential role in the quantification of $T_K$.
Moreover, multilevel effects can become interesting on their own (see Appendix).
A unique feature of NSN systems is that the bare Kondo couplings are strongly dependent on $\Delta/E_c$, which can be tuned by magnetic fields.
In SKE systems, the bare Kondo coupling is tipically smaller than in NSN systems because it contains a product of two small parameters: $\Gamma/\delta$ and $\delta/E_c$.

%Finally, we comment on the implications of having multichannel leads.
NNN systems can never have less than two Kondo channels, due to real spin degrees of freedom that are decoupled from the charge pseudospin.
In contrast, in NSN devices spin degrees of freedom are interwoven with the charge pseudospin (because the order parameter couples spin-up electrons with spin-down holes) and therefore the one-channel Kondo effect is in principle possible.
Unlike NU and SKE systems, most NNN and NSN systems carry wide junctions with multichannel leads.
However, multiple channels of a NSN system can be effectively mapped onto a one-channel Kondo model e.g. when the tunneling matrix amplitude is independent of energy.
Reaching a two-channel Kondo model in NNN systems with wide junctions is much less likely.

In sum, CKE in NSN systems ought to be experimentally measurable much like ordinary SKE is observable in semiconducting quantum dots, provided that single-channel Kondo models can be engineered. % for at least one of the NS junctions.

\section{Zero-Bias Conductance}
\label{sec:cond}

In previous sections we have described the charge-Kondo model that arises in mesoscopic NSN heterostrucures at sufficiently low energies.
In this section we investigate its fingerprints in the low-temperature electrical transport. 
We focus on the elastic contribution to the zero-bias conductance, which is dominant at $T<T^*$ because the superconducting gap freezes out inelastic quasiparticle excitations.    
Fingerprints of CKE can also get manifested in the low-temperature capacitance of superconducting islands coupled to a single normal metallic lead.\cite{zaikin1994,houzet2005}

The electrical conductance of mesoscopic NSN heterostructures has been widely discussed in the literature;\cite{zaikin1994,hekking1993,hergenrother1994,tinkham1995,falci2001,bignon2004,golubev2007,yeyati2007}  nonetheless, there appears to be no report on many-body anomalies in the low-temperature elastic conductance.
When bias voltages are weak, the current-voltage characteristics is encoded in    
\begin{equation}
\label{eq:i_vs_v}
I^\alpha=g_{\alpha\alpha'} V^\alpha \mbox{    } (\alpha\in\{L,R\}), 
\end{equation}
where $V^\alpha$ and $I^\alpha=\langle \hat{I}^\alpha\rangle=i e \,{\rm sgn}(\alpha)\langle[{\cal H}_{\rm eff}, \hat{N}^\alpha]\rangle$ are the voltage drop and the current across the $\alpha$-th junction, respectively.
$\hat{N}^{\alpha}$ denotes the electron number operator in the $\alpha$-th lead and ${\cal H}_{\rm eff}$ is given by Eq.~(\ref{eq:h_eff_ps}). Also, ${\rm sgn}(\alpha)\equiv 1(-1)$ for $\alpha=L (R)$.
 
The zero-bias conductance matrix in Eq.~(\ref{eq:i_vs_v}) can be expressed in terms of the Kubo formula:
\begin{equation}
\label{eq:kubo}
g_{\alpha\alpha'}=\lim_{\omega\to 0}\frac{1}{i\omega}\int_{0}^{\beta} d\tau e^{i\omega\tau}\langle {\cal T}_\tau\hat{I}^\alpha(\tau)\hat{I}^{\alpha'}(0)\rangle,
\end{equation}
where $\beta=1/(k_B T)$ and ${\cal T}_\tau$ is the time-ordering operator.
It is straightforward to obtain $\hat{I}^\alpha=\hat{I}^\alpha_{\rm single}+\hat{I}^\alpha_{\rm pair}$, where:
\begin{eqnarray}
\label{eq:currents}
\hat{I}^\alpha_{\rm single} &=& -i e\, {\rm sgn}(\alpha)\sum_{\alpha'\neq \alpha}\lambda_{||}^{\alpha\alpha'}\sum_{{\bf k},{\bf k}',\sigma}S^z c^\dagger_{\alpha{\bf k}\sigma} c_{{\alpha'\bf k}'\sigma}+{\rm h.c.}\nonumber\\
\hat{I}^\alpha_{\rm pair} &=& i e\, {\rm sgn}(\alpha)\sum_{\alpha'}\lambda_{\perp}^{\alpha\alpha'}\sum_{{\bf k},{\bf k}',\sigma} S^+ \sigma c_{\alpha{\bf k}\sigma} c_{\alpha'{\bf k}'-\sigma}+{\rm h.c.}\nonumber\\
\end{eqnarray}
$\langle\hat{I}_{\rm single}\rangle$ and $\langle\hat{I}_{\rm pair}\rangle$ are the single-particle and Cooper-pair tunneling contributions to the current, respectively.
$\alpha\neq\alpha'$ in the expression for $\hat{I}_{\rm single}$ simply recognizes the fact that local ($\alpha=\alpha'$) single-particle processes are normal reflections that do not contribute to the current. 
We treat the leads as non-superconducting ($\langle c_{\alpha {\bf k}\uparrow}c_{\alpha {\bf k}'\downarrow}\rangle=0$), non-magnetic ($\langle c^\dagger_{\alpha {\bf k}\uparrow}c_{\alpha {\bf k}'\downarrow}\rangle=0$) and infinitely long ($\langle c^\dagger_{\alpha {\bf k}\sigma} c_{\alpha {\bf k}'\sigma}\rangle\propto\delta_{{\bf k},{\bf k'}}$) conductors. 
With this in mind, the combination of Eqs.~(\ref{eq:kubo}) and (\ref{eq:currents}) yields
\begin{equation}
\label{eq:G_matrix}
g_{\alpha\alpha'}=\left(\begin{array}{cc} G_A^L+G_{EC}+G_{CA} & G_{EC}-G_{CA}\\G_{EC}-G_{CA} & G_A^R+G_{EC}+G_{CA}\end{array}\right),
\end{equation}
where
\begin{eqnarray}
\label{eq:conds}
G_A^{L (R)}&=& 2 G_0 \int d E (-\partial_E f) 2 [\nu\lambda_{\perp}^{LL (RR)}(E)]^2\nonumber\\
G_{CA}&=& G_0 \int d E (-\partial_E f) 2 [\nu\lambda_{\perp}^{LR}(E)]^2\nonumber\\
G_{EC}&=& G_0 \int d E (-\partial_E f) [\nu\lambda_{||}^{LR}(E)]^2,
\end{eqnarray}
$G_0=\pi e^2/\hbar$ and $f(E)=(\exp(\beta E)+1)^{-1}$.
%a factor of 4 smaller than Mahan's, b/c of the spin operators
$G_A^{L(R)}$ is the local Andreev conductance across the left (right) junction, $G_{CA}$ is the non-local or crossed Andreev conductance, and $G_{EC}$ is the elastic, single-particle co-tunneling conductance.\cite{falci2001,golubev2007}
The probability for an Andreev reflection process is proportional to $(\lambda_{x}^{\alpha\alpha'})^2+(\lambda^{\alpha\alpha'}_{y})^2=2(\lambda_\perp^{\alpha\alpha'})^2$, whereas the probability for a co-tunneling event is $\propto (\lambda^{LR}_{\rm ||})^2$.
The derivation of Eq.~(\ref{eq:G_matrix}) is standard\cite{mahan2000} except for the following two details.
First, $\lambda_{||,\perp}(E)$ is energy-dependent via the Kondo RG flow of Eq.~(\ref{eq:kt}), and diverges as $E\to 0$.
Second, the current operator in Eq.~(\ref{eq:currents}) contains $S^z$ and $S^\pm$; we have used $\langle {\cal T}_\tau S^z(\tau) S^z(0)\rangle=1/4$ and $\langle {\cal T}_\tau S^+(\tau) S^-(0)\rangle=1/2$ for $\tau>0$.

The matrix structure of $g_{\alpha\alpha'}$ in Eq.~(\ref{eq:G_matrix}) is simple to interpret. 
Regarding diagonal matrix elements,  each local Andreev reflection contributes twice as much as each co-tunneling or crossed Andreev reflection event because the former involves two electrons crossing the same junction while in the latter processes the two electrons cross different junctions. 
This fact is reflected by the extra factor of two in the first line of Eq.~(\ref{eq:conds}). 
Concerning the off-diagonal matrix elements, electron co-tunneling and crossed Andreev reflection contribute with opposite signs because in the later case it is a hole (rather than a electron) that tunnels out of the superconducting island.

Even though Eq.~(\ref{eq:G_matrix}) formally agrees with expressions derived in previous studies of NSN heterostructures,\cite{falci2001,bignon2004,golubev2007} there are conceptual novelties when the superconducting island behaves as a pseudospin 1/2 quantum impurity.
For instance, $G_{\rm EC}/G_{\rm CA}\to 1/2$ as the temperature is decreased, irrespective of the strength of electron-electron interactions in the superconducting island.
Therefore, crossed Andreev reflection and electron cotunneling do not cancel each other in the off-diagonal matrix elements of Eq.~(\ref{eq:G_matrix}). 
Other differences will become clear below.

In steady state $I_R=I_L\equiv I$ and the electrical response to the total voltage drop $V_R+V_L=V$ across the device can be evaluated by inverting Eq.~(\ref{eq:G_matrix}):
\begin{equation}
\label{eq:conductance}
G\equiv\frac{d I}{d V}=G_{EC}+\frac{(G_A^L+G_A^R) G_{CA}+G_A^L G_A^R}{G_A^L+G_A^R+4 G_{CA}}.
\end{equation}

Kondo correlations become apparent when we evaluate Eq.~(\ref{eq:conductance}) for $T<T^*$.
Let us suppose that the two NS junctions can be mapped onto a single-channel Kondo model. 
Combining Eqs.~(\ref{eq:lambda_coh}), (\ref{eq:Psi}) and ~(\ref{eq:conductance}) it follows that 
%$\lambda^{LR}_{||,\perp}=a \lambda^{LL}_{||,\perp}=a \lambda^{(1)}_{||,\perp}/(1+a^2)$ and $\lambda^{RR}_{||,\perp}=a^2 \lambda^{LL}_{||,\perp}=a^2 \lambda^{(1)}_{||,\perp}/(1+a^2)$, with $a=t_R/t_L$. 
\begin{equation}
\label{eq:cond_coherent}
G =\frac{G_0 t_L^2 t_R^2}{(t_L^2+t_R^2)^2}\int d E (-\partial_E f)\left((\nu \lambda_{||})^2+2(\nu \lambda_{\perp})^2\right),
\end{equation}
where $\lambda_{||,\perp}(E)$ obeys Eq.~(\ref{eq:kt}).

When $\lambda_{\perp}(T^*)=\lambda_{||}(T^*)$ (i.e. $E_c\simeq 0.5 \Delta$) and $T^*\gg T\gg T_K$, the integral in Eq.~(\ref{eq:cond_coherent}) yields
\begin{equation}
\label{eq:Glog}
G=G_0\frac{3 t_L^2 t_R^2}{4(t_L^2+t_R^2)^2}\frac{1}{\log^2(T/T_K)}.
\end{equation}
For $T<T_K$, the Kubo formula expression breaks down and the conductance converges to\cite{glazman2005} $G=2 (e^2/h)4 t_L^2 t_R^2/(t_L^2+t_R^2)^2$.
%\begin{equation}
%\label{eq:G_unitary_coh}
%G(T\to 0)=G_0\frac{t_L^2 t_R^2}{(t_L^2+t_R^2)^2}.
%\end{equation}
\begin{figure}[h]
\begin{center}
\includegraphics[scale=0.3, angle=270]{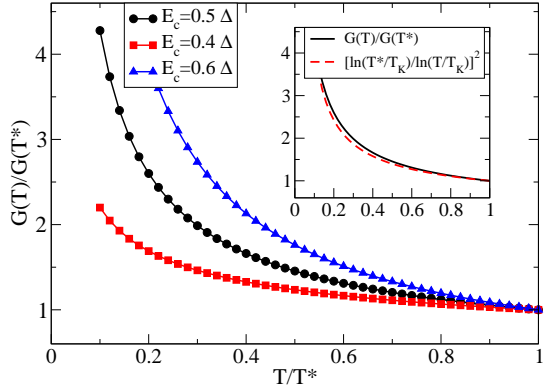}
\caption{Low-temperature (elastic) zero-bias conductance (Eq.~(\ref{eq:cond_coherent})) across the NSN heterostructure, in the single-channel Kondo model. 
The conductance is normalized to its value at $T=T^*$. Curves with red squares, black circles and blue triangles have $T_K=0.004 T^*$, $T_K\simeq 0.01 T^*$ and $T_K=0.05 T^*$, respectively. For simplicity we ignored $g_{\perp,||}$ in Eq.~(\ref{eq:l_T0}) and evaluated $T_K$ analytically neglecting third order terms in Eq.~(\ref{eq:kt}). We chose $\Delta=1000\delta$.  {\em Inset:} The solid line represents $G(T)/G(T^*)$ for $E_c=0.5\Delta$, as calculated from Eq.~(\ref{eq:cond_coherent}). Because $E_c=0.5\Delta$ yields the isotropic Kondo model ($\lambda_\perp(T^*)=\lambda_{||}(T^*)$), the solid line matches well with Eq.~(\ref{eq:Glog}) (dashed line).}
\label{fig:gcoh}
\end{center}
\end{figure}
The same results arise also in negative-U molecules with charge degeneracy,\cite{koch2007} in normal-metallic quantum dots with charge degeneracy\cite{glazman1990} and even in semiconducting quantum dots with spin degeneracy.\cite{glazman2005} 

The conductance for $\lambda_{\perp}(T^*)\neq \lambda_{||}(T^*)$ at $T^*\gg T\gg T_K$ is displayed in in Fig.~(\ref{fig:gcoh}).
$G(T)$ increases with $E_c/\Delta$, as expected from Fig.~(\ref{fig:fg}).
Since the anisotropy of the bare Kondo couplings disappears at sufficiently low temperatures, $G(T\ll T_K)$ is the same irrespective of $\lambda_{\perp}(T^*)/ \lambda_{||}(T^*)$.

The conductance curves of Fig.~(\ref{fig:gcoh}) have been neither predicted nor measured in previous studies of NSN systems.
For instance, Ref.~[\onlinecite{hekking1993}] neglected the inter-junction electron coherence (i.e. $G_{CA}=G_{EC}=0$ in Eq.~(\ref{eq:conductance})) and concluded that the zero-bias conductance through the superconducting grain will be independent of temperature, as long as $T\ll T^*$ and the gate voltage is tuned to the charge degeneracy point.
Contemporary experiments\cite{eiles1993,hergenrother1994} appeared to agree with this theoretical prediction, although the temperature dependence of the zero bias peak was not analyzed in detail. 
Yet according to our theory, the zero-bias conductance could display traces of a two-channel Kondo effect in islands with $L\gg\zeta$ .
This is the case in the experimental setup of Fig.~\ref{fig:setup}b, which should host a low-temperature enhancement of the Andreev reflection at the point contact, accompanied by a reduction of the Andreev reflection in the wide junction.  
%This behavior would become observable at $T\sim 10 {\rm mK}$.

%Due to the renormalization of the tip-grain coupling, the amplitude for Andreev reflection between the tip and the island grows at $T\ll T^*$(Eq.~(\ref{eq:Glog})) and accordingly the zero-bias peak in the tip-grain conductance becomes more pronounced.

%{\bf Watch out}
%{\bf inelastic tunneling contribution?}

\section{Charge-Kondo effect in the strong tunneling regime}
\label{sec:bozo}

Thus far all our calculations have concentrated on the weak tunneling regime, where $\Gamma\ll\delta$.
% the width $\Gamma$ of single-particle energy-levels in the superconducting island is much smaller than the average energy-level spacing $\delta$.
In this section we turn to highly transparent NS junctions ($\Gamma\gg\delta$) and determine the fate of CKE when charge in the superconducting island is strongly fluctuating. 
On one hand, in the weak tunneling regime the charge-Kondo temperature is enhanced by more transparent normal-superconducting junctions (Section~\ref{sec:disc}).
On the other hand, when charge in the dot fluctuates strongly it is no longer appropriate to regard the superconductor as a two-level system.
Which of these two conflicting trends prevails?
It turns out that the answer to this question depends on electron-electron interactions in the lead, which we have ignored up until now.
Strong tunneling and electron-electron interactions render bosonization as the most convenient method to employ in this section.
In particular we combine and suitably modify the approaches from Refs. ~[\onlinecite{glazman1999}] and [\onlinecite{affleck2000}].
% (see also Section 4 of Ref.~[\onlinecite{aleiner2002}] and references therein).

We consider a one-dimensional normal metallic wire connected to a superconductor (which need not be one-dimensional).
The lead occupies the negative half-axis ($x<0$), the NS interface is at $x=0$, and the superconductor is located at $x>0$.
We write the low-energy and long-wavelength Hamiltonian of this system as
\begin{equation}
\label{eq:1dmodel}
{\cal H}={\cal H}_{\rm lead}+{\cal H}_C+{\cal H}_{\rm B},
\end{equation}
where 
\begin{equation}
\label{eq:lead}
{\cal H}_{\rm lead}=i v_F\sum_\sigma\int_{-\infty}^{0} dx \left(\Psi^\dagger_{L\sigma} \partial_x \Psi_{L\sigma}-\Psi^\dagger_{R\sigma} \partial_x \Psi_{R\sigma}\right)
\end{equation}
is the Hamiltonian corresponding to the lead, 
\begin{equation}
\label{eq:charging}
{\cal H}_C=E_c\left[\int_{-\infty}^{0} dx \sum_\sigma \left(\Psi^\dagger_{L\sigma} \Psi_{L\sigma}+\Psi^\dagger_{R\sigma}\Psi_{R\sigma}\right)+N_g\right]^2 
\end{equation}
 is the charging energy and
\begin{equation}
\label{eq:boundary}
{\cal H}_{\rm B}= a V_B \sum_\sigma \Psi^\dagger_{\sigma}(0)\Psi_\sigma(0)+a \Delta_B [\Psi_{\uparrow}(0)\Psi_{\downarrow}(0)+{\rm h.c.}]
\end{equation}
is a boundary interaction ($a$ is the lattice spacing).
In Eqs.~(\ref{eq:lead}), (\ref{eq:charging}) and (\ref{eq:boundary}) we have expanded the original fermion operators as $\Psi_\sigma(x)\simeq e^{-i k_F x} \Psi_{L\sigma}(x)+e^{i k_F x} \Psi_{R\sigma}(x)$, where $k_F$ is the Fermi wavevector and $\Psi_{L\sigma}$ ($\Psi_{R\sigma})$ describes left-moving (right-moving) fermions. 
In Eq.~(\ref{eq:charging}) we have exploited total charge conservation in the dot-plus-lead system\cite{aleiner2002} so as to rewrite the charging energy in terms of the charge density in the lead (instead of in the dot).
Eq.~(\ref{eq:boundary}) follows from integrating out the superconductor at energy scales that are lower than the BCS gap $\Delta$. 
This integration results in a boundary term,\cite{affleck2000} which encodes normal scattering ($\propto V_B\ll\Delta$) as well as Andreev scattering ($\propto\Delta_B$).
Both $V_B$ and $\Delta_B$ (which can be taken to be purely real in our case) depend on microscopic details of the superconductor such as its Fermi velocity, $\Delta$ and $\delta$.
Also note that quasiparticle transmission into the superconductor is forbidden at energies below the gap.

We bosonize Eq.~(\ref{eq:1dmodel}) in the standard manner:\cite{giamarchi2003}
\begin{equation}
\Psi_{r\sigma}(x) =\frac{\eta_\sigma}{\sqrt{2\pi a}}e^{-\frac{i}{\sqrt{2}}\left[r\Phi_c(x)-\Theta_c(x)+r \sigma \Phi_s(x)-\sigma\Theta_s(x)\right]},
\end{equation}
where $\eta_\sigma$ is a Klein factor ($\eta_\uparrow\eta_\downarrow=-\eta_\downarrow\eta_\uparrow$, $\eta_\sigma \eta_\sigma=1$), $r=+$ $(-)$ for right- (left-) moving fermions and $\sigma=+ (-)$ for spin-up (-down) fermions.
$\Phi_c$ ($\Phi_s$) and $\Theta_c$ ($\Theta_s$) are the usual charge- (spin-) bosons.

It follows that
\begin{equation}
\label{eq:leadbozo}
{\cal H}_{\rm lead}=\frac{v_F}{2\pi}\sum_{\alpha\in\{c,s\}}\int_{-\infty}^{0} dx\left[K_\alpha(\partial_x\Theta_\alpha)^2+\frac{1}{K_\alpha}(\partial_x\Phi_\alpha)^2\right],
\end{equation}
where $K_c$ ($K_s$) is the Luttinger parameter in the charge (spin) sector ($K_s=1$ due to spin rotational symmetry).
Similarly,
\begin{equation}
\label{eq:chargebozo}
{\cal H}_C=E_c [\sqrt{2}\Phi_c(0)/\pi+N_g]^2,
\end{equation}
where $\sqrt{2}\Phi_c(0)/\pi$ equals the number of electrons contained in $x\in[-\infty,0]$ up to a constant (we take $\Phi_c(-\infty)=0$).

In order to bosonize ${\cal H}_B$, we use boundary conditions that are associated with perfect Andreev reflection at the NS interface:\cite{affleck2000} $\Psi_{R\uparrow}(0)=-i\Psi^\dagger_{L\downarrow}(0)$ and $\Psi_{R\downarrow}(0)=i\Psi^\dagger_{L\uparrow}(0)$. 
These boundary conditions provide a reasonable approximation for highly transparent NS interfaces with small Fermi surface mismatches. 
Furthermore, perfect Andreev reflection constitutes a renormalization group fixed point\cite{affleck2000} when $E_c=V_B=0$. 
In bosonic language, the Andreev boundary conditions can be translated as $\sqrt{2}\Theta_c(0)=\pi/2$ and $\sqrt{2}\Phi_s(0)=\pi$.
Consequently the boundary interaction at the Andreev fixed point reads
\begin{equation}
\label{eq:hb}
{\cal H}_B = -(2/\pi) V_B \cos[\sqrt{2}\Phi_c(0)],
\end{equation}
which describes single-particle backscattering (normal reflection) at the interface. 
Terms proportional to $\Delta_B$ do not appear in Eq.~(\ref{eq:hb}), which is consistent with the fact that boundary Andreev reflection is never a relevant perturbation at the Andreev fixed point.\cite{affleck2000}
%The second equality in Eq.~(\ref{eq:hb}) stems from the boundary conditions, which pin the spin boson at the interface.

Next, we determine how $E_c$ and $V_B$ evolve as the temperature is lowered.
Assuming that their bare values are small compared to $\Delta$, the leading order RG flow equations read 
\begin{equation}
\frac{d E_c}{d l}= E_c \mbox{   ;   } \frac{d V_B}{dl}=(1-K_c) V_B,
\end{equation}
where $l=\ln(\Delta/T)$. 
Thus $E_c$ is a relevant perturbation, and $V_B$ is relevant, marginal or irrelevant depending on whether electron-electron interactions in the lead are repulsive ($K_c<1$), non-interacting ($K_c=1$) or attractive ($K_c>1$), respectively.
This is markedly different from the non-superconducting system analyzed in Ref.~[\onlinecite{glazman1999}], where $d V_B/d l=(1-K_c/2) V_B$ and backscattering is a relevant perturbation for non-interacting electrons.
The underlying reason for this qualitative difference is that a perfectly transmitted channel at a NN interface has a conductance of $2 e^2/h$, whereas a perfectly transmitted channel at a NS interface has a conductance of $4 e^2/h$ (due to perfect Andreev reflection).
%The scaling dimension of $V_B$ (which equals $1-K_c$) can be derived in the standard way\cite{gogolin} (but keep in mind that the interacting term exists only at the boundary.)

Since either $E_c$ or $V_B$ is certain to grow under RG, the Andreev fixed point is unstable.
The nature of the new fixed point depends crucially on whether $E_c$ or $V_B$ reaches strong coupling first.
If $E_c$ wins,  the charge boson at the boundary gets pinned to a value dictated by the gate voltage, $\sqrt{2}\Phi_c(0)/\pi\to-N_g$, and the ground state is non-degenerate.
Since the gate charge $N_g$ is a continuous variable, the discreteness of charge is unimportant in this case and the system is nowhere equivalent to a two-level system. 

If $V_B$ prevails instead, then the charge boson at the interface becomes pinned through $\sqrt{2}\Phi_c(0)/\pi\to 2n$, for any integer $n$.
The discreteness of charge becomes important through the discreteness of $n$, although there are an infinite number of degenerate ground states.
This energy degeneracy between different values of $n$ is lifted by ${\cal H}_C$ (Eq.~(\ref{eq:chargebozo})).
Choosing $N_g$ appropriately (e.g. $N_g=1$) the lowest energy state becomes two-fold degenerate ($n=0$ and $n=-1$), where the two states differ from one another by {\em two} electrons.
The remaining minima of the cosine potential are at energies $\gtrsim E_c$ higher, where $E_c$ is the renormalized charging energy.
Therefore in the strong coupling limit of $V_B$ the ground state of Eq.~(\ref{eq:1dmodel}) behaves as a two-level quantum system. 
Rare tunneling events between the $n=0$ and $n=-1$ charge states (induced by the kinetic energy terms in Eq.~(\ref{eq:leadbozo})) are tantamount to the rare spin-flip events of a magnetic impurity that is weakly coupled to a bath of itinerant fermions. 
Those tunneling events become prominent as energy is lowered further, and ultimately culminate in the charge-Kondo effect. 

In sum, for CKE to emerge in the strong tunneling regime it is necessary that $V_B$ dominate over $E_c$ under renormalization group. 
In terms of the parameters of the original Hamiltonian, this condition reads
\begin{equation}
\label{eq:cke_co}
\frac{\Delta}{E_c(0)}\gg\left(\frac{\Delta}{V_B (0)}\right)^\frac{1}{1-K_c} \mbox{   ;    } K_c<1,
\end{equation}
where $V_B(0)$ and $E_c(0)$ are the values of the backscattering potential and the charging energy at energy scales of order $\Delta$. 
In mesoscopic NSN devices Eq.~(\ref{eq:cke_co}) is difficult to satisfy unless electron-electron interactions in the lead are strongly repulsive and/or the bare charging energy is very small.
The temperature at which $V_B$ reaches the strong coupling regime is $T_0\equiv\Delta e^{-l_c}$, where $l_c$ is defined through  $V_B(l_c)\simeq\Delta$. 
Thus
\begin{equation}
T_0\simeq \Delta(V_B(0)/\Delta)^\frac{1}{1-K_c}\ll\Delta.
\end{equation} 
%$T_0$ can be interpreted as the temperature below which the ground state Eq.~(\ref{eq:1dmodel}) maps onto a pseudospin 1/2 system.
If $T_0\gg E_c(l_c)=E_c(0) e^{l_c}$, inelastic charge excitations (i.e. transitions to $n\neq 0,-1$) are present at temperatures $T_0>T>E_c(l_c)$. 
In this case, by analogy with Ref.~[\onlinecite{glazman1999}], the dimensionless Kondo coupling is $\nu\lambda\sim E_c(l_c)/T_0$ and the Kondo temperature scales like $T_K\sim E_c(l_c) \exp(-1/\nu\lambda)\ll E_c(l_c)\ll T_0$.
If $T_0<E_c(l_c)$, inelastic excitations are forbidden as soon as $T<T_0$ and $T_K$ could be a sizeable fraction of $T_0$.
Yet because $T_0$ is typically very small in weakly interacting leads ($K_c\simeq 1$), $T_K$ is unlikely to be detectable.
Prospects might be better in leads made from carbon nanotubes, where\cite{kane1997} $K_c\simeq 0.2\ll 1$

In Section~\ref{sec:disc} we learned that the charge-Kondo temperature is unobservably low when $\Gamma/\delta\to 0$, though it increases rapidly as $\Gamma/\delta$ grows.
In this section we have seen that $T_K$ is very small when $\Gamma/\delta\gg 1$ as well, at least for Fermi liquid leads. 
Therefore, intermediate tunneling strengths are most suitable for the observability of CKE in NSN devices.

\section{Summary and conclusions}
\label{sec:conc}

Partly motivated by the recent experimental search for charge-Kondo correlations in Tl-doped PbTe, we have developed a theory for the charge- Kondo effect in artificially fabricated, chaotic superconducting islands that are weakly connected to non-superconducting leads. 
We have focused on tunable superconducting charge-qubits where the energy gap in the excitation spectrum is larger than the electrostatic charging energy. 
The low-energy transport and thermodynamic properties of these systems showcase Kondo-like behavior when the electrostatic energies of two charge states differing by a Cooper pair become degenerate. 
A single-channel anisotropic Kondo model is found to describe superconducting grains of size smaller than the superconducting coherence length, which are coupled to normal conductors either through atomic point contacts or semiconducting nanowires.
The same model could also apply for wider metal-superconducting contacts as long as the tunneling amplitude is independent of energy in a narrow strip around the Fermi energy. 
For this model the Kondo temperature can reach $\simeq 10 {\rm mK}$ with aluminum grains and $\simeq 20 {\rm mK}$ with niobium grains, which is comparable to the temperature of the cryostat in current NSN devices.
Higher Kondo temperatures might be accessible by engineering normal-superconductor contacts of intermediate transparency.  
Our proposal for the charge-Kondo effect differs qualitatively from previous proposals in non-superconducting single-electron devices and in valence-skipping compounds, and may be easier to observe.

Avenues for further theoretical research in NSN devices include understanding the transition from single-channel to multichannel Kondo models, examining the crossover from weak to strong dot-lead tunneling, characterizing the charge-Kondo effect when the normal-superconductor junctions are disordered, and searching for a low-temperature many-body enhancement of the tunneling magnetoresistance when the leads are magnetized. 
% and interference in the normal-superconducting junctions.\cite{hekking1994}
%Other open problems include understanding the charge Kondo effect away from equilibrium, as well as studying capacitively coupled superconducting islands from the viewpoint of two-impurity Kondo models.
%It will also be interesting to consider ferromagnetic (or magnetized) leads and to search for a low-temperature enhancement of the tunneling magnetoresistance\cite{giazotto2006,das2008} due to the charge-Kondo effect.   

\acknowledgements
I am indebted to I. Affleck for many helpful discussions. 
This research has been funded through a Junior Fellowship of the Canadian Institute for Advanced Research.

\begin{widetext}

\appendix

\section{``First-Principles'' Derivation of CKE in NSN Devices}

In Section~\ref{sec:kondo} we introduced a low energy effective Kondo model, Eq.~(\ref{eq:h_eff_ps}), from which we inferred the Kondo effect by simply invoking standard renormalization group equations.
While we provided physical arguments to justify the emergence of such a model at temperatures lower than $T^*$, we did not present a rigorous derivation.
The main objective of this Appendix is to validate Eq.~(\ref{eq:h_eff_ps}) by deriving hallmark Kondo-like divergences from the full microscopic (or ``first-principles'') Hamiltonian, Eq.~(\ref{eq:model}).

The procedure we follow has been introduced elsewhere.\cite{garate2011}
The task undertaken here consists of evaluating the amplitude for Andreev reflection to fourth order in tunneling.
Andreev reflection transforms an itinerant electron into an itinerant hole, while adding a Cooper pair to the superconducting island. 
This can be interpreted as a process that flips the pseudospins of the superconducting island as well as of the incident fermion.
The initial and final states for this process are
\begin{equation}
|i\rangle=c^\dagger_{k\uparrow}|\o\rangle|2M\rangle \mbox{   ;   } |f\rangle= c_{k\downarrow}|\o\rangle|2M+2\rangle,
\end{equation}
which are eigenstates of Eq.~(\ref{eq:model}) in absence of tunneling.
Tunneling induces $|i\rangle\to|f\rangle$ transitions, and the transition amplitude $A_{i\to f}$ may be computed from certain matrix elements of an effective Hamiltonian derived perturbatively from Eq.~(\ref{eq:model}).\cite{garate2011}
One arrives at $A_{i\to f} = A^{(2)}_{i\to f}+A^{(4)}_{i\to f}+...$, where
\begin{eqnarray}
A^{(2)}_{i\to f}&=&\sum_{n}\frac{\langle f|{\cal H}_T|n\rangle\langle n|{\cal H}_T|i\rangle}{E_i-E_n}\nonumber\\
A^{(4)}_{i\to f} &=& \sum_{n_1,n_2,n_3}\frac{\langle f|{\cal H}_T|n_1\rangle\langle n_1|{\cal H}_T|n_2\rangle\langle n_2|{\cal H}_T|n_3\rangle\langle n_3|{\cal H}_T|i\rangle}{(E_i-E_{n_1})(E_i-E_{n_2})(E_i-E_{n_3})}-\epsilon_2\sum_n\frac{\langle f|{\cal H}_T|n\rangle\langle n|{\cal H}_T|i\rangle}{(E_i-E_n)^2}-A^{(2)}_{i\to f} \sum_n\frac{|\langle f|{\cal H}_T|n\rangle|^2}{(E_i-E_n)^2}\nonumber
\end{eqnarray}
and $|n\rangle, |n_1\rangle, ...$ denote intermediate states of energies $E_n$, $E_{n_1}$...in absence of tunneling.
In addition,
\begin{equation}
\label{eq:e2}
\epsilon_2=\sum_n\frac{|\langle i|{\cal H}_T|n\rangle|^2}{E_n-E_i}= t^2\sum_n\sum_q \Theta(\xi_q)\left(\frac{1}{E_c-\xi_q-\sqrt{\epsilon_n^2+\Delta^2}}-\frac{1}{3E_c+\xi_q+\sqrt{\epsilon_n^2+\Delta^2}}\right),
\end{equation}
where in the second equality we have assumed particle-hole symmetry and energy-independent tunneling amplitudes. 
$\Theta(x)$ is the step function.

$\nu A^{(2)}_{i\to f}$ is identical to $(\Gamma/\delta)f_\perp$ of Section~\ref{sec:kondo}, and was first computed in Ref.~[\onlinecite{hekking1993}]. 
It is finite and non-singular for $E_c<\Delta$.
Yet $A^{(4)}_{i\to f}$, which has not been previously computed, hosts infrared (IR) logarithmic divergences for any $E_c/\Delta$.
These IR divergences signal the onset of Kondo correlations.

Evaluating $A^{(4)}_{i\to f}$ is straightforward in principle but cumbersome in practice.
For convenience we attach Tables \ref{tab:a1}-\ref{tab:a4}, which collect all possible intermediate state configurations for the first term in the expression of $A^{(4)}_{i\to f}$.
Many of the individual amplitudes entering Tables \ref{tab:a1}-\ref{tab:a4} are ultraviolet (UV) divergent in the infinite-bandwidth limit.
Remarkably, all UV divergences in $A^{(4)}_{i\to f}$ end up cancelling one another after summing over all the amplitudes.
This indicates that details of the energy spectrum at high energies are not important for the Kondo effect, and endows universality to results derived in this paper. 
In fact, the bulk contribution to $A_{i\to f}$ originates from states with $|\epsilon_n|\lesssim\Delta$.

Following a numerical sum of all the amplitudes, we arrive at
\begin{equation}
\label{eq:a4}
\nu A_{i\to f}=\frac{\Gamma}{\delta}f_\perp+2\frac{\Gamma^2}{\delta^2} f_\perp f_{||}\ln\frac{\Delta}{\omega}+\frac{\Gamma^2}{\delta^2}g_\perp,
\end{equation}
where $g_\perp$ is a dimensionless function of $E_c/\Delta$.
To the present order approximation, Eq.~(\ref{eq:a4}) can be recasted as
\begin{equation}
\label{eq:a4b}
\nu A_{i\to f}\simeq \nu\lambda_\perp+2(\nu\lambda_\perp)(\nu\lambda_{||})\ln\frac{T^*}{\omega},
\end{equation}
where 
\begin{equation}
\label{eq:l_T}
\nu\lambda_\perp = \frac{\Gamma}{\delta} f_{\perp}\left[1+\frac{\Gamma}{\delta} \frac{g_\perp}{f_\perp}+ 2\frac{\Gamma}{\delta} f_{||}\ln\frac{\Delta}{T^*}\right]\mbox{  and      } \,\,\nu\lambda_{||} = \frac{\Gamma}{\delta} f_{||}\left[1+\frac{\Gamma}{\delta} \frac{g_{||}}{f_{||}}+2\frac{\Gamma}{\delta}\frac{f_\perp^2}{f_{||}}\ln\frac{\Delta}{T^*}\right]
\end{equation}
 are the values of the Kondo couplings at the energy scale $T^*$ (see Eq.~(\ref{eq:l_T0})) and we have neglected $O(\Gamma^3)$ terms.
Eq.~(\ref{eq:a4b}) provides a first-principles proof for the second line of Eq.~(\ref{eq:kt}) and establishes the anisotropic Kondo model as an appropriate low energy theory for Eq.~(\ref{eq:model}).
Admittedly, the energy scale $T^*$ does not appear naturally in our zero-temperature calculation.
 Rather we introduce it in hindsight (i.e.  with the knowledge that only at energies below $T^*$ can the superconducting island behave as a two-level system) and arrange the rest of the terms accordingly.
The logarithmic terms in Eq.~(\ref{eq:l_T}) indicate that the Kondo couplings begin to renormalize starting at energy scales of order $\Delta$.
%, which may be substantially larger than $T^*$.
$g_\perp$ is plotted in Fig.~(\ref{fig:fg}), where it is shown that 4th order tunneling events can either enhance (if $g_\perp>0$) or deplete (if $g_\perp<0$) the Andreev reflection amplitude, depending on $E_c/\Delta$.
We have not evaluated $g_{||}$, which would require additional Tables of intermediate state configurations.
We expect $g_{||}\simeq g_\perp$ for $E_c\simeq 0.5\Delta$.
Our perturbative analysis becomes insufficient when $(\Gamma/\delta) |g_\perp|\sim 1$, which occurs at $E_c\simeq 0.8 \Delta$ for $\Gamma/\delta\simeq 0.1$.
Of course, the breakdown of the perturbative regime does not preclude a Kondo effect; it just means that we cannot compute the Kondo temperature reliably using the weak-tunneling approach.

\begin{table*}[h]
\caption{Virtual elastic processes to fourth order in single-particle tunneling.
$n$, $m$ and $q$ are dummy variables to be summed over. 
The amplitudes labelled with a number are either UV and/or IR divergent.
The amplitudes labelled with a letter are non-divergent.
The amplitudes grouped with the same number or letter are identical to one another in presence of particle-hole symmetry.
For each of these groups, half of the configurations involve electrons ($c^\dagger_q$) and half involve holes ($c_q$).
For simplicity we have assumed the tunneling amplitude to be simply a constant; however, our results can be readily generalized to include more complicated situations.
In the derivation of Fig.~(\ref{fig:fg}) we use $\sum_q\to \nu\int d\xi$ and $\sum_n\to (1/\delta)\int dn$.
N.B.: in these Tables (and in these Tables only) $E_n$ stands for $\sqrt{\epsilon_n^2+\Delta^2}$.}
% (do not confuse it with the energy of an intermediate state configuration).}
\vspace{0.1 in}
\begin{tabular}{c c c c c }
\hline\hline
Label & $|n_1\rangle$ & $|n_2\rangle$ & $|n_3\rangle$ & Contribution to $A^{(4)}_{i\to f}$ (in units of $t^4$)\\
\hline\hline
1 & $|\o\rangle \gamma^\dagger_{n\uparrow}|2M+2\rangle$ & $c^\dagger_{q\sigma}|\o\rangle \gamma^\dagger_{n\uparrow} \gamma^\dagger_{m,-\sigma} |2M+2\rangle$   & $|\o\rangle \gamma^\dagger_{n\uparrow} |2M+2\rangle$   & $2\frac{u_n v_n v_m^2 \Theta(\xi_q)}{(E_c-E_n)^2(\xi_q+E_n+E_m)}$\\[1ex]
\hline
1 &  $|\o\rangle \gamma^\dagger_{n\uparrow}|2M+2\rangle$ & $c_{q\sigma}|\o\rangle \gamma^\dagger_{n\uparrow} \gamma^\dagger_{m\sigma}|2M+2\rangle$ & $|\o\rangle \gamma^\dagger_{n\uparrow} |2M+2\rangle$ & $2 \frac{u_n v_n u_m^2 \Theta(-\xi_q)}{(E_c-E_n)^2 (-\xi_q+E_n+E_m)}$\\[1ex]
\hline
8 & $|\o\rangle \gamma^\dagger_{n\uparrow}|2M+2\rangle$ & $c^\dagger_{k\uparrow}|\o\rangle \gamma^\dagger_{n\uparrow} \gamma^\dagger_{m\downarrow}|2M+2\rangle$ & $c^\dagger_{k\uparrow} c_{q\uparrow}|\o\rangle \gamma^\dagger_{n\uparrow} |2M+2\rangle $ & $\frac{u_n v_n v_m^2 \Theta(-\xi_q)}{(E_c-E_n)(E_n+E_m)(E_c+\xi_q-E_n)}$ \\[1ex]
\hline
7 & $|\o\rangle \gamma^\dagger_{n\uparrow}|2M+2\rangle$ & $c^\dagger_{k\uparrow}|\o\rangle \gamma^\dagger_{n\uparrow} \gamma^\dagger_{m\downarrow} |2M+2\rangle$ & $c^\dagger_{k\uparrow} c_{q\downarrow}|\o\rangle \gamma^\dagger_{m\downarrow}|2M+2\rangle$ & $\frac{u_m v_m v_n^2 \Theta(-\xi_q)}{(E_c-E_n)(E_n+E_m)(E_c+\xi_q-E_m)}$ \\[1ex]
\hline
2 & $|\o\rangle \gamma^\dagger_{n\uparrow}|2M+2\rangle$ & $c^\dagger_{q\downarrow}|\o\rangle \gamma^\dagger_{n\uparrow} \gamma^\dagger_{m\uparrow}|2M+2\rangle$ & $|\o\rangle \gamma^\dagger_{m\uparrow} |2M+2\rangle$ & $-\frac{u_m v_m v_n^2 \Theta(\xi_q)}{(E_c-E_n)(\xi_q+E_n+E_m)(E_c-E_m)}$\\[1ex]
\hline
4 & $|\o\rangle \gamma^\dagger_{n\uparrow}|2M+2\rangle$ & $c_{q\downarrow}|\o\rangle|2M+2\rangle$ & $|\o\rangle \gamma^\dagger_{m\uparrow}|2M+2\rangle$ & $-\frac{u_m v_m v_n^2 \Theta(-\xi_q)}{(E_c-E_n) \xi_q (E_c-E_m)}$\\[1ex]
\hline
11  & $|\o\rangle \gamma^\dagger_{n\uparrow}|2M+2\rangle$ & $c_{q\downarrow}|\o\rangle|2M+2\rangle$ & $c^\dagger_{k\uparrow} c_{q\downarrow}|\o\rangle \gamma^\dagger_{m\downarrow}|2M+2\rangle$ & $-\frac{u_m v_m v_n^2 \Theta(-\xi_q)}{(E_c-E_n)\xi_q (E_c+\xi_q-E_m)}$\\[1ex]
\hline
3 & $|\o\rangle \gamma^\dagger_{n\uparrow}|2M+2\rangle$ & $c^\dagger_{q\uparrow}|\o\rangle |2M\rangle$ & $c^\dagger_{q\uparrow} c^\dagger_{k\uparrow}|\o\rangle \gamma^\dagger_{m\downarrow}|2M\rangle$ & $\frac{u_n v_n v_m^2 \Theta(\xi_q)}{(E_c-E_n) \xi_q (\xi_q+3 E_c+E_m)}$\\[1ex]
\hline
4 & $|\o\rangle \gamma^\dagger_{n\uparrow}|2M+2\rangle$ & $c^\dagger_{q\uparrow}|\o\rangle |2M\rangle$ & $|\o\rangle \gamma^\dagger_{m\uparrow}|2M\rangle$ & $\frac{u_n v_n u_m^2 \Theta(\xi_q)}{(E_c-E_n)\xi_q (E_c-E_m)}$\\[1ex]
\hline
A & $|\o\rangle \gamma^\dagger_{n\uparrow}|2M+2\rangle$ & $c^\dagger_{q\sigma}|\o\rangle \gamma^\dagger_{n\uparrow} \gamma^\dagger_{m,-\sigma}|2M+2\rangle$ & $c^\dagger_{q\sigma} c^\dagger_{k\uparrow}|\o\rangle \gamma^\dagger_{m,-\sigma}|2M\rangle$ & $-2\frac{u_n v_n v_m^2 \Theta(\xi_q)}{(E_c-E_n)(\xi_q+E_n+E_m)(\xi_q+3 E_c+E_m)}$\\[1ex]
\hline
B & $|\o\rangle \gamma^\dagger_{n\uparrow}|2M+2\rangle$ & $c_{q\sigma}|\o\rangle \gamma^\dagger_{n\uparrow} \gamma^\dagger_{m\sigma}|2M+2\rangle$ & $c^\dagger_{k\uparrow} c_{q\sigma}|\o\rangle \gamma^\dagger_{m\sigma}|2M\rangle$ & $-2\frac{u_n v_n u_m^2 \Theta(-\xi_q)}{(E_c-E_n)(\xi_q-E_n-E_m)(\xi_q+E_c-E_m)}$\\[1ex]
\hline
5 & $|\o\rangle \gamma^\dagger_{n\uparrow}|2M+2\rangle$ & $c^\dagger_{k\uparrow}|\o\rangle \gamma^\dagger_{n\uparrow} \gamma^\dagger_{m\downarrow}|2M+2\rangle$ & $c^\dagger_{k\uparrow} c^\dagger_{q\downarrow}|\o\rangle \gamma^\dagger_{n\uparrow}|2M\rangle$ & $\frac{u_m v_m v_n^2 \Theta(\xi_q)}{(E_c-E_n)(E_n+E_m)(\xi_q+3 E_c+E_n)}$ \\[1ex]
\hline
6 & $|\o\rangle \gamma^\dagger_{n\uparrow}|2M+2\rangle$ & $c^\dagger_{k\uparrow}|\o\rangle \gamma^\dagger_{n\uparrow} \gamma^\dagger_{m\downarrow}|2M+2\rangle$ & $c^\dagger_{k\uparrow} c^\dagger_{q\uparrow}|\o\rangle \gamma^\dagger_{m\downarrow}|2M\rangle$ & $\frac{u_n v_n v_m^2 \Theta(\xi_q)}{(E_c-E_n)(E_n+E_m)(\xi_q+3 E_c+E_m)}$\\[1ex]
\hline
22 & $|\o\rangle \gamma^\dagger_{n\uparrow}|2M+2\rangle$ & $c^\dagger_{q\downarrow}|\o\rangle \gamma^\dagger_{n\uparrow} \gamma^\dagger_{m\uparrow}|2M+2\rangle$ & $c^\dagger_{q\downarrow} c^\dagger_{k\uparrow}|\o\rangle \gamma^\dagger_{n\uparrow}|2M\rangle$ & $\frac{u_m v_m v_n^2 \Theta(\xi_q)}{(E_c-E_n)(\xi_q+E_n+E_m)(\xi_q+3 E_c+E_n)}$\\[1ex]
\hline
10 & $|\o\rangle \gamma^\dagger_{n\uparrow}|2M+2\rangle$ & $c_{q\uparrow}|\o\rangle \gamma^\dagger_{n\uparrow} \gamma^\dagger_{m\uparrow}|2M+2\rangle$ & $ c_{q\uparrow} c^\dagger_{k\uparrow}|\o\rangle \gamma^\dagger_{n\uparrow}|2M+2\rangle$ & $\frac{u_n v_n u_m^2 \Theta(-\xi_q)}{(E_c-E_n)(\xi_q-E_n-E_m)(E_c+\xi_q-E_n)}$\\[1ex]
\hline
2 & $|\o\rangle \gamma^\dagger_{n\uparrow}|2M+2\rangle$ & $c_{q\uparrow}|\o\rangle \gamma^\dagger_{n\uparrow} \gamma^\dagger_{m\uparrow}|2M+2\rangle$ & $|\o\rangle \gamma^\dagger_{m\uparrow}|2 M\rangle$ & $\frac{u_n v_n u_m^2 \Theta(-\xi_q)}{(E_c-E_n)(\xi_q-E_n-E_m)(E_c-E_m)}$\\[1ex]
\hline\hline 
E & $c_{k'\downarrow} c^\dagger_{q\sigma}|\o\rangle \gamma^\dagger_{n,-\sigma}|2M+2\rangle$ & $c^\dagger_{q\sigma}|\o\rangle \gamma^\dagger_{m\uparrow} \gamma^\dagger_{n,-\sigma}|2M+2\rangle$ & $|\o\rangle \gamma^\dagger_{m\uparrow}|2M+2\rangle$ & $-2\frac{u_m v_m v_n^2 \Theta(\xi_q)}{(E_c-\xi_q-E_n)(-\xi_q-E_n-E_m)(E_c-E_m)}$ \\[1ex]
\hline
8 & $c_{k'\downarrow} c^\dagger_{q\downarrow}|\o\rangle \gamma^\dagger_{n\uparrow}|2M+2\rangle$ & $c_{k'\downarrow}|\o\rangle \gamma^\dagger_{m\downarrow} \gamma^\dagger_{n\uparrow} |2M+2\rangle$ & $|\o\rangle \gamma^\dagger_{n\uparrow}|2M+2\rangle$ & $-\frac{u_n v_n u_m^2 \Theta(\xi_q)}{(E_c-\xi_q-E_n)(-E_m-E_n)(E_c-E_n)}$\\[1ex]
\hline
B & $c_{k'\downarrow} c^\dagger_{q\sigma}|\o\rangle \gamma^\dagger_{n,-\sigma}|2M+2\rangle$ & $c^\dagger_{k\uparrow} c_{k'\downarrow} c^\dagger_{q\sigma}|\o\rangle \gamma^\dagger_{n,-\sigma} \gamma^\dagger_{m\downarrow}|2M+2\rangle$ & $c^\dagger_{k\uparrow} c_{k'\downarrow} |\o\rangle \gamma^\dagger_{m\downarrow} |2M+2\rangle$ & $-2\frac{u_m v_m v_n^2 \Theta(\xi_q)}{(E_c-\xi_q-E_n)(-\xi_q-E_n-E_m)(E_c-E_m)}$ \\[1ex]
\hline
9 & $c_{k'\downarrow} c^\dagger_{q\uparrow}|\o\rangle \gamma^\dagger_{n\downarrow}|2M+2\rangle$ & $c_{k'\downarrow}|\o\rangle \gamma^\dagger_{m\uparrow} \gamma^\dagger_{n\downarrow} |2M+2\rangle$ & $c_{k'\downarrow} c^\dagger_{k\uparrow}|\o\rangle \gamma^\dagger_{n\downarrow}|2M+2\rangle$ & $-\frac{u_n v_n u_m^2\Theta(\xi_q)}{(E_c-\xi_q-E_n)(-E_n-E_m)(E_c-E_n)}$\\[1ex]
\hline
10 & $c_{k'\downarrow} c^\dagger_{q\downarrow}|\o\rangle \gamma^\dagger_{n\uparrow}|2M+2\rangle$ & $c^\dagger_{q\downarrow}|\o\rangle \gamma^\dagger_{m\uparrow} \gamma^\dagger_{n\uparrow}|2M+2\rangle$ & $|\o\rangle \gamma^\dagger_{n\uparrow}|2M+2\rangle$ & $\frac{u_n v_n v_m^2 \Theta(\xi_q)}{(E_c-\xi_q-E_n)(-\xi_q-E_n-E_m)(E_c-E_n)}$ \\[1ex]
\hline
11 & $c_{k'\downarrow} c^\dagger_{q\uparrow}|\o\rangle \gamma^\dagger_{n\downarrow}|2M+2\rangle$ & $c^\dagger_{q\uparrow}|\o\rangle|2M\rangle$ & $|\o\rangle \gamma^\dagger_{m\uparrow}|2M+2\rangle$ & $-\frac{u_n v_n u_m^2 \Theta(\xi_q)}{(E_c-\xi_q-E_n)(-\xi_q)(E_c-E_m)}$ \\[1ex]
\hline
11 & $c_{k'\downarrow} c^\dagger_{q\downarrow}|\o\rangle \gamma^\dagger_{n\uparrow}|2M+2\rangle$ & $c^\dagger_{k\uparrow} c_{k'\downarrow} c^\dagger_{q\downarrow}|\o\rangle |2M\rangle$ & $c^\dagger_{k\uparrow} c_{k'\downarrow}|\o\rangle \gamma^\dagger_{m\downarrow}|2M+2\rangle$ & $-\frac{u_n v_n u_m^2 \Theta(\xi_q)}{(E_c-\xi_q-E_n)(-\xi_q)(E_c-E_m)}$\\[1ex]
\hline
12 & $c_{k'\downarrow} c^\dagger_{q\uparrow}|\o\rangle \gamma^\dagger_{n\downarrow}|2M+2\rangle$ & $c^\dagger_{q\uparrow}|\o\rangle |2M\rangle$ & $c^\dagger_{q\uparrow} c^\dagger_{k\uparrow}|\o\rangle \gamma^\dagger_{m\downarrow}|2M\rangle$ & $\frac{u_n v_n v_m^2 \Theta(\xi_q)}{(E_c-\xi_q-E_n)(-\xi_q)(-\xi_q-3 E_c-E_m)}$\\[1ex]
\hline
13 & $c_{k'\downarrow} c^\dagger_{q\sigma}|\o\rangle \gamma^\dagger_{n,-\sigma}|2M+2\rangle$ & $c^\dagger_{q\sigma}|\o\rangle \gamma^\dagger_{m\uparrow} \gamma^\dagger_{n,-\sigma}|2M+2\rangle$ & $c^\dagger_{q\sigma} c^\dagger_{k\uparrow}|\o\rangle \gamma^\dagger_{n,-\sigma}|2 M\rangle$ & $-2 \frac{u_m v_m v_n^2 \Theta(\xi_q)}{(E_c-\xi_q-E_n)(-\xi_q-E_n-E_m)(-\xi_q-3 E_c-E_n)}$\\[1ex]
\hline
C & $c_{k'\downarrow} c^\dagger_{q\downarrow}|\o\rangle \gamma^\dagger_{n\uparrow}|2M+2\rangle$ & $c^\dagger_{q\downarrow}|\o\rangle \gamma^\dagger_{m\uparrow} \gamma^\dagger_{n\uparrow}|2M+2\rangle$ & $c^\dagger_{q\downarrow} c^\dagger_{k\uparrow}|\o\rangle \gamma^\dagger_{m\uparrow}|2 M\rangle$ & $\frac{u_n v_n v_m^2 \Theta(\xi_q)}{(E_c-\xi_q-E_n)(-\xi_q-E_n-E_m)(-\xi_q-3 E_c-E_m)}$ \\[1ex]
\hline
14 & $c_{k'\downarrow} c^\dagger_{q\downarrow}|\o\rangle \gamma^\dagger_{n\uparrow}|2M+2\rangle$ & $c^\dagger_{k\uparrow} c_{k'\downarrow} c^\dagger_{q\downarrow}|\o\rangle|2M\rangle$ & $c^\dagger_{k\uparrow} c^\dagger_{q\downarrow}|\o\rangle \gamma^\dagger_{m\uparrow}|2M\rangle$ & $\frac{u_n v_n v_m^2 \Theta(\xi_q)}{(E_c-\xi_q-E_n)(-\xi_q)(-\xi_q-3 E_c-E_m)}$ \\[1ex]
\hline
15 & $c_{k'\downarrow} c^\dagger_{q\uparrow}|\o\rangle \gamma^\dagger_{n\downarrow}|2M+2\rangle$ & $c^\dagger_{k\uparrow} c_{k'\downarrow} c^\dagger_{q\uparrow}|\o\rangle \gamma^\dagger_{n\downarrow} \gamma^\dagger_{m\downarrow}|2M+2\rangle$ & $c^\dagger_{k\uparrow} c_{k'\downarrow}|\o\rangle \gamma^\dagger_{n\downarrow}|2M+2\rangle$ & $-\frac{u_n v_n v_m^2 \Theta(\xi_q)}{(E_c-\xi_q-E_n)(\xi_q+E_n+E_m)(E_c-E_n)}$\\[1ex]
\hline
13 & $c_{k'\downarrow} c^\dagger_{q\downarrow}|\o\rangle \gamma^\dagger_{n\uparrow}|2M+2\rangle$ & $c^\dagger_{k\uparrow} c_{k'\downarrow} c^\dagger_{q\downarrow}|\o\rangle \gamma^\dagger_{n\uparrow} \gamma^\dagger_{m\downarrow}|2M+2\rangle$ & $c^\dagger_{k\uparrow} c^\dagger_{q\downarrow}|\o\rangle \gamma^\dagger_{n\uparrow}|2M\rangle$ & $-\frac{u_m v_m v_n^2 \Theta(\xi_q)}{(E_c-\xi_q-E_n)(-\xi_q-E_n-E_m)(-\xi_q-3 E_c-E_n)}$\\[1ex]
\hline
13 & $c_{k'\downarrow} c^\dagger_{q\uparrow}|\o\rangle \gamma^\dagger_{n\downarrow}|2M+2\rangle$ & $c^\dagger_{k\uparrow} c_{k'\downarrow} c^\dagger_{q\uparrow}|\o\rangle \gamma^\dagger_{n\downarrow} \gamma^\dagger_{m\downarrow}|2M+2\rangle$ & $c^\dagger_{k\uparrow} c^\dagger_{q\uparrow}|\o\rangle \gamma^\dagger_{n\downarrow}|2M\rangle$ & $-\frac{u_m v_m v_n^2 \Theta(\xi_q)}{(E_c-\xi_q-E_n)(-\xi_q-E_n-E_m)(-\xi_q-3 E_c-E_n)}$\\[1ex]
\hline
C & $c_{k'\downarrow} c^\dagger_{q\uparrow}|\o\rangle \gamma^\dagger_{n\downarrow}|2M+2\rangle$ & $c^\dagger_{k\uparrow} c_{k'\downarrow} c^\dagger_{q\uparrow}|\o\rangle \gamma^\dagger_{n\downarrow} \gamma^\dagger_{m\downarrow}|2M+2\rangle$ & $c^\dagger_{k\uparrow} c^\dagger_{q\uparrow}|\o\rangle \gamma^\dagger_{m\downarrow}|2M\rangle$ & $\frac{u_n v_n v_m^2 \Theta(\xi_q)}{(E_c-\xi_q-E_n)(-\xi_q-E_n-E_m)(-\xi_q-3 E_c-E_m)}$ \\[1ex]
\hline
7& $c_{k'\downarrow} c^\dagger_{q\downarrow}|\o\rangle \gamma^\dagger_{n\uparrow}|2M+2\rangle$ & $c_{k'\downarrow}|\o\rangle \gamma^\dagger_{n\uparrow} \gamma^\dagger_{m\downarrow}|2M+2\rangle$ & $c_{k'\downarrow} c^\dagger_{k\uparrow}|\o\rangle \gamma^\dagger_{m\downarrow}|2M\rangle$ & $-\frac{u_n v_n u_m^2 \Theta(\xi_q)}{(E_c-\xi_q-E_n)(-E_n-E_m)(E_c-E_m)}$\\[1ex]
\hline
7& $c_{k'\downarrow} c^\dagger_{q\uparrow}|\o\rangle \gamma^\dagger_{n\downarrow}|2M+2\rangle$ & $c_{k'\downarrow}|\o\rangle \gamma^\dagger_{n\downarrow} \gamma^\dagger_{m\uparrow}|2M+2\rangle$ & $|\o\rangle \gamma^\dagger_{m\uparrow}|2M\rangle$ & $-\frac{u_n v_n u_m^2 \Theta(\xi_q)}{(E_c-\xi_q-E_n)(-E_n-E_m)(E_c-E_m)}$\\[1ex]
\hline\hline
\label{tab:a1}
\end{tabular}
\end{table*}

\begin{table*}[h]
\caption{Continuation of Table \ref{tab:a1}}
\vspace{0.1 in}
\begin{tabular}{c c c c c }
\hline\hline
Label & $|n_1\rangle$ & $|n_2\rangle$ & $|n_3\rangle$ & Contribution to $A^{(4)}_{i\to f}$ (in units of $t^4$)\\
\hline\hline
16 & $c_{k'\downarrow} c^\dagger_{k\uparrow}|\o\rangle \gamma^\dagger_{n\downarrow}|2M+2\rangle$ & $c_{k'\downarrow} c^\dagger_{k\uparrow} c^\dagger_{q\sigma}|\o\rangle \gamma^\dagger_{m,-\sigma} \gamma^\dagger_{n\downarrow}|2M+2\rangle$ & $c_{k'\downarrow} c^\dagger_{k\uparrow}|\o\rangle \gamma^\dagger_{n\downarrow}|2M+2\rangle$ & $2\frac{u_n v_n v_m^2 \Theta(\xi_q)}{(E_c-E_n)^2(\xi_q+E_n+E_m)}$  \\[1ex]
\hline
16 & $c_{k'\downarrow} c^\dagger_{k\uparrow}|\o\rangle \gamma^\dagger_{n\downarrow}|2M+2\rangle$ & $c_{k'\downarrow} c^\dagger_{k\uparrow} c_{q\sigma} |\o\rangle \gamma^\dagger_{m\sigma} \gamma^\dagger_{n\downarrow} |2M+2\rangle$ & $c_{k'\downarrow} c^\dagger_{k\uparrow}|\o\rangle \gamma^\dagger_{n\downarrow}|2M+2\rangle$ & $-2\frac{u_n v_n u_m^2 \Theta(-\xi_q)}{(E_c-E_n)^2 (\xi_q-E_n-E_m)}$\\[1ex]
\hline
7 & $c_{k'\downarrow} c^\dagger_{k\uparrow}|\o\rangle \gamma^\dagger_{n\downarrow}|2M+2\rangle$ & $c^\dagger_{k\uparrow}|\o\rangle \gamma^\dagger_{n\downarrow} \gamma^\dagger_{m\uparrow} |2M+2\rangle$ & $c^\dagger_{k\uparrow} c_{q\uparrow} |\o\rangle \gamma^\dagger_{m\uparrow}|2M+2\rangle$ & $\frac{u_m v_m v_n^2 \Theta(-\xi_q)}{(E_c-E_n)(E_n+E_m)(\xi_q+E_c-E_m)}$\\[1ex]
\hline
17 & $c_{k'\downarrow} c^\dagger_{k\uparrow}|\o\rangle \gamma^\dagger_{n\downarrow}|2M+2\rangle$ & $c_{k'\downarrow} c^\dagger_{k\uparrow} c^\dagger_{q\uparrow}|\o\rangle \gamma^\dagger_{m\downarrow} \gamma^\dagger_{n\downarrow}|2M+2\rangle$ & $c_{k'\downarrow} c^\dagger_{k\uparrow}|\o\rangle \gamma^\dagger_{m\downarrow}|2M+2\rangle$ & $-\frac{u_m v_m v_n^2 \Theta(\xi_q)}{(E_c-E_n)(\xi_q+E_n+E_m)(E_c-E_m)}$\\[1ex]
\hline
18 & $c_{k'\downarrow} c^\dagger_{k\uparrow}|\o\rangle \gamma^\dagger_{n\downarrow}|2M+2\rangle$ & $c_{k'\downarrow} c^\dagger_{k\uparrow} c_{q\uparrow}|\o\rangle |2M+2\rangle$ & $c_{k'\downarrow} c^\dagger_{k\uparrow}|\o\rangle \gamma^\dagger_{m\downarrow}|2M+2\rangle$ & $-\frac{u_m v_m v_n^2 \Theta(-\xi_q)}{(E_c-E_n)\xi_q (E_c-E_m)}$ \\[1ex]
\hline
11 & $c_{k'\downarrow} c^\dagger_{k\uparrow}|\o\rangle \gamma^\dagger_{n\downarrow}|2M+2\rangle$ & $c_{k'\downarrow} c^\dagger_{k\uparrow} c_{q\uparrow}|\o\rangle|2M+2\rangle$ & $c^\dagger_{k\uparrow} c_{q\uparrow}|\o\rangle \gamma^\dagger_{m\uparrow}|2M+2\rangle$ & $-\frac{u_m v_m v_n^2 \Theta(-\xi_q)}{(E_c-E_n)\xi_q(E_c+\xi_q-E_m)}$ \\[1ex]
\hline
9 & $c_{k'\downarrow} c^\dagger_{k\uparrow}|\o\rangle \gamma^\dagger_{n\downarrow}|2M+2\rangle$ & $c^\dagger_{k\uparrow}|\o\rangle \gamma^\dagger_{n\downarrow} \gamma^\dagger_{m\uparrow}|2M+2\rangle$ & $c^\dagger_{k\uparrow} c_{q\downarrow}|\o\rangle \gamma^\dagger_{n\downarrow}|2M+2\rangle$ & $\frac{u_n v_n v_m^2 \Theta(-\xi_q)}{(E_c-E_n)(E_n+E_m)(E_c+\xi_q-E_n)}$ \\[1ex]
\hline
18 & $c_{k'\downarrow} c^\dagger_{k\uparrow}|\o\rangle \gamma^\dagger_{n\downarrow}|2M+2\rangle$ & $c_{k'\downarrow} c^\dagger_{k\uparrow} c^\dagger_{q\downarrow}|\o\rangle|2M\rangle$ & $c_{k'\downarrow} c^\dagger_{k\uparrow}|\o\rangle \gamma^\dagger_{m\downarrow}|2M+2\rangle$ & $\frac{u_n v_n u_m^2 \Theta(\xi_q)}{(E_c-E_n)\xi_q(E_c-E_m)}$\\[1ex]
\hline
19 & $c_{k'\downarrow} c^\dagger_{k\uparrow}|\o\rangle \gamma^\dagger_{n\downarrow}|2M+2\rangle$ & $c_{k'\downarrow} c^\dagger_{k\uparrow} c^\dagger_{q\downarrow}|\o\rangle|2M\rangle$ & $c^\dagger_{k\uparrow} c^\dagger_{q\downarrow}|\o\rangle \gamma^\dagger_{m\uparrow}|2M\rangle$ & $\frac{u_n v_n v_m^2 \Theta(\xi_q)}{(E_c-E_n)\xi_q(\xi_q+3 E_c+ E_m)}$\\[1ex]
\hline
20 & $c_{k'\downarrow} c^\dagger_{k\uparrow}|\o\rangle \gamma^\dagger_{n\downarrow}|2M+2\rangle$ & $c_{k'\downarrow} c^\dagger_{k\uparrow} c^\dagger_{q\uparrow}|\o\rangle \gamma^\dagger_{m\downarrow} \gamma^\dagger_{n\downarrow}|2M+2\rangle$ & $c^\dagger_{k\uparrow} c^\dagger_{q\uparrow}|\o\rangle \gamma^\dagger_{n\downarrow} |2 M\rangle$ & $\frac{u_m v_m v_n^2 \Theta(\xi_q)}{(E_c-E_n)(\xi_q+E_n+E_m)(\xi_q+3 E_c+E_n)}$\\[1ex]
\hline
E & $c_{k'\downarrow} c^\dagger_{k\uparrow}|\o\rangle \gamma^\dagger_{n\downarrow}|2M+2\rangle$ & $c_{k'\downarrow} c^\dagger_{k\uparrow} c_{q\sigma}|\o\rangle \gamma^\dagger_{m\sigma} \gamma^\dagger_{n\downarrow}|2M+2\rangle$ & $c^\dagger_{k\uparrow} c_{q\sigma}|\o\rangle \gamma^\dagger_{m\sigma}|2M\rangle$ & $-2\frac{u_n v_n u_m^2 \Theta(-\xi_q)}{(E_c-E_n)(\xi_q-E_m-E_n)(E_c+\xi_q-E_m)}$\\[1ex]
\hline
5 & $c_{k'\downarrow} c^\dagger_{k\uparrow}|\o\rangle \gamma^\dagger_{n\downarrow}|2M+2\rangle$ & $c^\dagger_{k\uparrow}|\o\rangle \gamma^\dagger_{n\downarrow} \gamma^\dagger_{m\uparrow} |2M+2\rangle$ & $c^\dagger_{k\uparrow} c^\dagger_{q\uparrow}|\o\rangle \gamma^\dagger_{n\downarrow}|2M\rangle$ & $\frac{u_m v_m v_n^2 \Theta(\xi_q)}{(E_c-E_n)(E_n+E_m)(\xi_q+3 E_c+E_n)}$\\[1ex]
\hline
D & $c_{k'\downarrow} c^\dagger_{k\uparrow}|\o\rangle \gamma^\dagger_{n\downarrow}|2M+2\rangle$ & $c_{k'\downarrow} c^\dagger_{k\uparrow} c^\dagger_{q\sigma}|\o\rangle \gamma^\dagger_{m,-\sigma} \gamma^\dagger_{n\downarrow}|2M+2\rangle$ & $c^\dagger_{k\uparrow} c^\dagger_{q\sigma}|\o\rangle \gamma^\dagger_{m,-\sigma}|2M\rangle$ & $-2\frac{u_n v_n v_m^2 \Theta(\xi_q)}{(E_c-E_n)(\xi_q+E_n+E_m)(\xi_q+3 E_c+E_m)}$\\[1ex]  
\hline
17 & $c_{k'\downarrow} c^\dagger_{k\uparrow}|\o\rangle \gamma^\dagger_{n\downarrow}|2M+2\rangle$ & $c_{k'\downarrow} c^\dagger_{k\uparrow} c_{q\downarrow}|\o\rangle \gamma^\dagger_{m\downarrow} \gamma^\dagger_{n\downarrow}|2M+2\rangle$ & $c_{k'\downarrow} c^\dagger_{k\uparrow}|\o\rangle \gamma^\dagger_{m\downarrow}|2M\rangle$ & $\frac{u_n v_n u_m^2\Theta(-\xi_q)}{(E_c-E_n)(\xi_q-E_n-E_m)(E_c-E_m)}$\\[1ex]
\hline
15 & $c_{k'\downarrow} c^\dagger_{k\uparrow}|\o\rangle \gamma^\dagger_{n\downarrow}|2M+2\rangle$ & $c_{k'\downarrow} c^\dagger_{k\uparrow} c_{q\downarrow}|\o\rangle \gamma^\dagger_{m\downarrow} \gamma^\dagger_{n\downarrow}|2M+2\rangle$ & $c^\dagger_{k\uparrow} c_{q\downarrow}|\o\rangle \gamma^\dagger_{n\downarrow}|2M+2\rangle$ & $\frac{u_n v_n u_m^2 \Theta(-\xi_q)}{(E_c-E_n)(\xi_q-E_n-E_m)(E_c+\xi_q-E_n)}$\\[1ex]
\hline
21 & $c_{k'\downarrow} c^\dagger_{k\uparrow}|\o\rangle \gamma^\dagger_{n\downarrow}|2M+2\rangle$ & $c^\dagger_{k\uparrow}|\o\rangle \gamma^\dagger_{n\downarrow} \gamma^\dagger_{m\uparrow}|2M+2\rangle$ & $c^\dagger_{k\uparrow} c^\dagger_{q\downarrow}|\o\rangle \gamma^\dagger_{m\uparrow}|2M\rangle$ & $\frac{u_n v_n v_m^2 \Theta(\xi_q)}{(E_c-E_n)(-E_n-E_m)(-3 E_c-\xi_q-E_m)}$\\[1ex]
\hline\hline
D & $c_{k'\downarrow} c_{q\sigma}|\o\rangle \gamma^\dagger_{n\sigma}|2M+2\rangle$ & $c_{q\sigma}|\o\rangle \gamma^\dagger_{n\sigma} \gamma^\dagger_{m\uparrow}|2M+2\rangle$ & $|\o\rangle \gamma^\dagger_{m\uparrow} |2M+2\rangle$ & $-2\frac{u_m v_m u_n^2 \Theta(-\xi_q)}{(\xi_q-3 E_c-E_n)(\xi_q-E_n-E_m)(E_c-E_m)}$ \\[1ex]
\hline
A & $c_{k'\downarrow} c_{q\sigma}|\o\rangle \gamma^\dagger_{n\sigma}|2M+2\rangle$ & $c_{k'\downarrow} c_{q\sigma} c^\dagger_{k\uparrow}|\o\rangle \gamma^\dagger_{n\sigma} \gamma^\dagger_{m\downarrow}|2M+2\rangle$ & $c_{k'\downarrow} c^\dagger_{k\uparrow}|\o\rangle \gamma^\dagger_{m\downarrow}|2M+2\rangle$ & $-2\frac{u_m v_m u_n^2 \Theta(-\xi_q)}{(\xi_q-3 E_c-E_n)(\xi_q-E_n-E_m)(E_c-E_m)}$\\[1ex]
\hline
3 & $c_{k'\downarrow} c_{q\downarrow}|\o\rangle \gamma^\dagger_{n\downarrow}|2M+2\rangle$ & $c_{q\downarrow}|\o\rangle|2M+2\rangle$ & $|\o\rangle \gamma^\dagger_{m\uparrow}|2M+2\rangle$ & $\frac{u_m v_m u_n^2 \Theta(-\xi_q)}{(\xi_q-3 E_c-E_n) \xi_q (E_c-E_m)}$\\[1ex]
\hline
14 & $c_{k'\downarrow} c_{q\downarrow}|\o\rangle \gamma^\dagger_{n\downarrow}|2M+2\rangle$ & $c_{q\downarrow}|\o\rangle|2M+2\rangle$ & $c^\dagger_{k\uparrow} c_{q\downarrow}|\o\rangle \gamma^\dagger_{m\downarrow}|2M+2\rangle$ & $\frac{u_m v_m u_n^2 \Theta(-\xi_q)}{(\xi_q-3 E_c-E_n) \xi_q (\xi_q+E_c-E_m)}$\\[1ex]
\hline
19 & $c_{k'\downarrow} c_{q\uparrow}|\o\rangle \gamma^\dagger_{n\uparrow}|2M+2\rangle$ & $c_{k'\downarrow} c_{q\uparrow} c^\dagger_{k\uparrow}|\o\rangle |2M+2\rangle$ & $c_{k'\downarrow} c^\dagger_{k\uparrow}|\o\rangle \gamma^\dagger_{m\downarrow}|2M+2\rangle$ & $\frac{u_m v_m u_n^2 \Theta(-\xi_q)}{(\xi_q-3 E_c- E_n)\xi_q (E_c-E_m)}$\\[1ex]
\hline
12 & $c_{k'\downarrow} c_{q\uparrow}|\o\rangle \gamma^\dagger_{n\uparrow}|2M+2\rangle$ & $c_{k'\downarrow} c_{q\uparrow} c^\dagger_{k\uparrow}|\o\rangle|2M+2\rangle$ & $c_{q\uparrow} c^\dagger_{k\uparrow}|\o\rangle \gamma^\dagger_{m\uparrow}|2M+2\rangle$ & $ \frac{u_m v_m u_n^2 \Theta(-\xi_q)}{(\xi_q-3 E_c- E_n) \xi_q (\xi_q+ E_c-E_m)}$\\[1ex]
\hline
13 & $c_{k'\downarrow} c_{q\sigma}|\o\rangle \gamma^\dagger_{n\sigma}|2M+2\rangle$ & $c_{q\sigma}|\o\rangle \gamma^\dagger_{n\sigma} \gamma^\dagger_{m\uparrow}|2M+2\rangle$ & $c^\dagger_{k\uparrow} c_{q\sigma}|\o\rangle \gamma^\dagger_{n\sigma}|2 M\rangle$ & $-2\frac{u_m v_m u_n^2 \Theta(-\xi_q)}{(\xi_q- 3 E_c-E_n)(\xi_q-E_n-E_m)(\xi_q+E_c-E_n)}$\\[1ex]
\hline
13 & $c_{k'\downarrow} c_{q\sigma}|\o\rangle \gamma^\dagger_{n\sigma}|2M+2\rangle$ & $c_{k'\downarrow} c_{q\sigma} c^\dagger_{k\uparrow}|\o\rangle \gamma^\dagger_{n\sigma} \gamma^\dagger_{m\downarrow}|2M+2\rangle$ & $c_{q\sigma} c^\dagger_{k\uparrow}|\o\rangle \gamma^\dagger_{n\sigma}|2M\rangle$ & $-2\frac{u_m v_m u_n^2 \Theta(-\xi_q)}{(\xi_q- 3 E_c- E_n)(\xi_q- E_n- E_m)(\xi_q+E_c-E_n)}$\\[1ex]
\hline
5 & $c_{k'\downarrow} c_{q\downarrow}|\o\rangle \gamma^\dagger_{n\downarrow}|2M+2\rangle$ & $c_{k'\downarrow}|\o\rangle \gamma^\dagger_{m\uparrow} \gamma^\dagger_{n\downarrow}|2M+2\rangle$ & $c_{k'\downarrow} c^\dagger_{k\uparrow}|\o\rangle \gamma^\dagger_{n\downarrow}|2 M\rangle$ & $\frac{u_m v_m u_n^2 \Theta(-\xi_q)}{(\xi_q-3 E_c-E_n)(-E_n-E_m)(E_c-E_n)}$\\[1ex]
\hline
20 & $c_{k'\downarrow} c_{q\uparrow}|\o\rangle \gamma^\dagger_{n\uparrow}|2M+2\rangle$ & $c_{q\uparrow}|\o\rangle \gamma^\dagger_{n\uparrow} \gamma^\dagger_{m\uparrow}|2M+2\rangle$ & $|\o\rangle \gamma^\dagger_{n\uparrow}|2M\rangle$ & $\frac{u_m v_m u_n^2 \Theta(-\xi_q)}{(\xi_q-3 E_c-E_n)(\xi_q-E_n-E_m)(E_c-E_n)}$ \\[1ex]
\hline
22 & $c_{k'\downarrow} c_{q\downarrow}|\o\rangle \gamma^\dagger_{n\downarrow}|2M+2\rangle$ & $c_{k'\downarrow} c_{q\downarrow} c^\dagger_{k\uparrow}|\o\rangle \gamma^\dagger_{n\downarrow} \gamma^\dagger_{m\downarrow}|2M+2\rangle$ & $c_{k'\downarrow} c^\dagger_{k\uparrow}|\o\rangle \gamma^\dagger_{n\downarrow}|2M\rangle$ & $\frac{u_m v_m u_n^2 \Theta(-\xi_q)}{(\xi_q-3 E_c-E_n)(\xi_q-E_n- E_m)(E_c-E_n)}$\\[1ex]
\hline
5 & $c_{k'\downarrow} c_{q\uparrow}|\o\rangle \gamma^\dagger_{n\uparrow}|2M+2\rangle$ & $c_{k'\downarrow}|\o\rangle \gamma^\dagger_{m\downarrow} \gamma^\dagger_{n\uparrow}|2M+2\rangle$ & $|\o\rangle \gamma^\dagger_{n\uparrow}|2M\rangle$ & $\frac{u_m v_m u_n^2 \Theta(-\xi_q)}{(\xi_q-3 E_c-E_n)(-E_n-E_m)(E_c-E_n)}$\\[1ex]
\hline
6 & $c_{k'\downarrow} c_{q\uparrow}|\o\rangle \gamma^\dagger_{n\uparrow}|2M+2\rangle$ & $c_{k'\downarrow}|\o\rangle \gamma^\dagger_{m\downarrow} \gamma^\dagger_{n\uparrow}|2M+2\rangle$ & $c_{k'\downarrow} c^\dagger_{k\uparrow}|\o\rangle \gamma^\dagger_{m\downarrow}|2M+2\rangle$ & $\frac{u_m v_m u_n^2 \Theta(-\xi_q)}{(\xi_q-3 E_c-E_n)(-E_n-E_m)(E_c-E_m)}$\\[1ex]
\hline
21 & $c_{k'\downarrow} c_{q\downarrow}|\o\rangle \gamma^\dagger_{n\downarrow}|2M+2\rangle$ & $c_{k'\downarrow}|\o\rangle \gamma^\dagger_{m\uparrow} \gamma^\dagger_{n\downarrow}|2M+2\rangle$ & $|\o\rangle \gamma^\dagger_{m\uparrow}|2M+2\rangle$ & $\frac{u_m v_m u_n^2 \Theta(-\xi_q)}{(\xi_q-3 E_c-E_n)(-E_n-E_m)(E_c-E_m)}$\\[1ex]
\hline
C& $c_{k'\downarrow} c_{q\uparrow}|\o\rangle \gamma^\dagger_{n\uparrow}|2M+2\rangle$ & $c_{q\uparrow}|\o\rangle \gamma^\dagger_{n\uparrow} \gamma^\dagger_{m\uparrow}|2M+2\rangle$ & $c^\dagger_{k\uparrow} c_{q\uparrow}|\o\rangle \gamma^\dagger_{m\uparrow}|2M+2\rangle$ & $\frac{u_m v_m u_n^2 \Theta(-\xi_q)}{(\xi_q-3 E_c-E_n)(\xi_q-E_n-E_m)(\xi_q+E_c-E_m)}$\\[1ex]
\hline
C& $c_{k'\downarrow} c_{q\downarrow}|\o\rangle \gamma^\dagger_{n\downarrow}|2M+2\rangle$ & $c_{k'\downarrow} c_{q\downarrow} c^\dagger_{k\uparrow}|\o\rangle \gamma^\dagger_{n\downarrow} \gamma^\dagger_{m\downarrow}|2M+2\rangle$ & $c^\dagger_{k\uparrow} c_{q\downarrow}|\o\rangle \gamma^\dagger_{m\downarrow}|2M+2\rangle$ & $\frac{u_m v_m u_n^2 \Theta(-\xi_q)}{(\xi_q-3 E_c-E_n)(\xi_q-E_n-E_m)(\xi_q+E_c-E_m)}$\\[1ex]
\hline\hline
\label{tab:a4}
\end{tabular}
\end{table*}

\end{widetext}

\end{document}